\documentclass{jpp}
\usepackage{graphicx}

\usepackage[utf8]{inputenc}
\usepackage[T1]{fontenc}
\usepackage{amsmath}
\usepackage{xcolor}
\usepackage{bigints}
\usepackage{dsfont}
\usepackage[colorlinks=true]{hyperref}

\DeclareMathOperator*{\argmin}{arg\,min}

\shorttitle{Adjoint ideal ballooning}
\shortauthor{Gaur et al.}

\title{An adjoint-based method for optimizing MHD equilibria against the infinite-$n$, ideal ballooning mode}

\author{Rahul Gaur\aff{1}
  \corresp{\email{rgaur@umd.edu}},
  Stefan Buller\aff{1}, Maximilian E. Ruth\aff{2}, Matt Landreman\aff{1}, Ian G. Abel\aff{1}, \and William D. Dorland\aff{1, 3}}
  
\affiliation{\aff{1} Institute for Research in Electronics and Applied Physics, University of Maryland, College Park, 20740, MD, USA
\aff{2}Center for Applied Mathematics, Cornell University, Ithaca, 14850, NY, USA
\aff{3} Department of Physics, University of Maryland, College Park, 20740, MD, USA}

\begin{document}

\maketitle

\begin{abstract}
We demonstrate a fast adjoint-based method to optimize tokamak and stellarator equilibria against a pressure-driven instability known as the infinite-$n$ ideal ballooning mode. We present three finite-$\beta$ (the ratio of thermal to magnetic pressure) equilibria: one tokamak equilibrium and two stellarator equilibria that are unstable against the ballooning mode. Using the self-adjoint property of ideal MHD, we construct a technique to rapidly calculate the change in the growth rate, a measure of ideal ballooning instability. Using the~\texttt{SIMSOPT} framework, we then implement our fast adjoint gradient-based optimizer to minimize the growth rate and find stable equilibria for each of the three originally unstable equilibria.
\end{abstract}

\section{Introduction}
Magnetic confinement is currently considered the most promising way~\citep{baalrud2020community} to achieve the United States' goal of building a pilot fusion power plant that generates net electricity before 2040~\citep{national2021bringing}. Most advanced fusion reactor designs today are based on two main designs that use magnetic confinement: tokamaks and stellarators. These devices work by using strong magnetic fields to keep a hot, dense plasma at their center. The main difference between tokamaks and stellarators lies in their geometric shape. Tokamaks are symmetric about a fixed axis, whereas stellarators are not. Due to the difference in axisymmetry, the tokamak equilibria are 2D-axisymmetric, and the stellarator equilibria are 3D.

For a fixed magnetic field strength, the power density $P$ of a fusion device scales as $\beta^2$, where $\beta$ is the ratio of the plasma pressure to the magnetic pressure. Since current tokamaks and stellarators are low-$\beta$ devices, one way to improve the efficiency of these devices is to increase the operating $\beta$. However, doing so creates a large pressure gradient from the center to the edge of the device, which is a source of a variety of pressure-driven, magnetohydrodynamic (MHD) and kinetic instabilities. One of the important pressure-driven instabilities is the infinite-$n$ ideal ballooning mode. 

 In tokamaks, there have been numerous studies that have used the ideal ballooning mode to determine the plasma beta limit. This led to the development of codes such as~\texttt{EPED}~\citep{snyder2007stability} and~\texttt{ELITE}~\citep{ELITE} that can determine edge pressure profiles with reasonable accuracy.
In stellarators, the ideal ballooning mode might not cause a disruption, but~\citet{tang1980kinetic} have shown that it is directly related to a kinetic instability known as the Kinetic Ballooning Mode (KBM). Recent articles have shown the detrimental effects of KBM turbulence on stellarators for finite beta values~\citep{aleynikova2018kinetic, mckinney2021kinetic}. The KBM is similar to the ideal ballooning mode, albeit with additional kinetic effects. Therefore, the ideal ballooning mode could be used as a proxy for KBM stability.

Numerous studies have been conducted to optimize tokamaks~\citep{miller1979shape, bernard1981systematic} and stellarators~\citep{sanchez2000ballooning, gates2017recent} against ideal MHD instabilities. However, such calculations can be computationally costly and time consuming. In this paper, we use the self-adjoint property of ideal MHD to devise an adjoint-based method that speeds up the optimization of 2D and 3D equilibria against the infinite-$n$, ideal ballooning mode. Adjoint-based methods have been extensively used for aeronautical design~\citep{giles2000introduction} and recently in the context of stellarator optimization (see \citet{paul2021gradient} and references therein). Using this technique, we can speed up the process of finding equilibria that are stable against the ballooning mode.

The remainder of this paper is divided as follows: in~\S\ref{sec:equilibrium}, we briefly describe the fundamentals of a general 3D MHD equilibrium and follow it with the details of the \texttt{VMEC} equilibrium solver~\citep{hirshman1983steepest_VMEC} in~\S\ref{subsec:VMEC}. Using \texttt{VMEC}, we obtain and present the details of one 2D-axisymmetric equilibrium in~\S\ref{subsec:2D-axisym-eqbm}  and two 3D equilibria in~\S\ref{subsec:3D-NCSX-eqbm} and~\S\ref{subsec:3D-HBERG-eqbm}. In~\S\ref{sec:ideal-balloning mode}, we present the physical, mathematical, and numerical details used to solve the infinite-$n$, ideal ballooning equation. We then analyze the susceptibility of the chosen local equilibria to the ideal ballooning instability. In~\S\ref{subsec:iball-self-adjoint}, we explain the self-adjoint property of the ideal ballooning equation. We also explain how the ballooning eigenvalue can be used as a proxy to stabilize the equilibria against the KBM. Using the self-adjoint technique, we formulate an adjoint method which we explain and test in~\S\ref{sec:Adjoint-method}. In~\S\ref{sec:optimizn-details}, we present the details of the overall optimization process and present an adjoint-based optimizer using the \texttt{SIMSOPT}~\citep{landreman2021simsopt} framework. In the penultimate section, we present our results, comparing the optimized stable equilibria with their initial, unstable counterparts. Finally, in~\S\ref{sec:summary-and-conclusions} we summarize our work and discuss possible ways in which it can be extended.

\section{Ideal MHD equilibrium}
\label{sec:equilibrium}
In this section, we start with the general form of a three-dimensional, divergence-free magnetic field. We use this form to represent the magnetic field in tokamaks and stellarators. After that, we briefly describe the steady-state, ideal MHD, force-balance equation. In~\S\ref{subsec:VMEC}, we explain the details of solving the ideal MHD force-balance equation using the \texttt{VMEC} code. Finally, we present the details of three MHD equilibria in~\S\ref{subsec:2D-axisym-eqbm},~\S\ref{subsec:3D-NCSX-eqbm}, and~\S\ref{subsec:3D-HBERG-eqbm} that we will use throughout this study.

A divergence-free magnetic field $\boldsymbol{B}$ can be written in the Clebsch form~\citep{d2012flux}
\begin{equation}
    \boldsymbol{B} = \bnabla\alpha_{\mathrm{t}} \times \bnabla \psi_{\mathrm{p}}.
    \label{eqn:Div-free-B1}
\end{equation}
The form~\eqref{eqn:Div-free-B1} is generally used for tokamak equilibria. For stellarators, we use the following equivalent representation
\begin{equation}
    \boldsymbol{B} =   \bnabla \psi \times \bnabla\alpha_{\mathrm{s}}.
    \label{eqn:Div-free-B2}
\end{equation}
We will focus on solutions whose magnetic field lines lie on closed nested toroidal surfaces, known as flux surfaces. For tokamaks, we label the flux surfaces with their enclosed poloidal flux $\psi_{\mathrm{p}}$ whereas for stellarators, we use the enclosed toroidal flux $\psi$. On each flux surface, lines of constant $\alpha_{\mathrm{t}}$ and $\alpha_{\mathrm{s}}$ coincide with the magnetic field-lines in tokamaks and stellartors, respectively. Thus, the variables $\alpha_{\mathrm{t}}$ and $\alpha_{\mathrm{s}}$ are known as field line labels.

To facilitate the calculation of various physical quantities from a general equilibrium solver, we use multiple coordinate systems. We will use the right-handed cylindrical coordinate system $(R, \zeta, Z)$ where $R$ and $Z$ are the radial and vertical distances from the origin and $\zeta$ is the azimuthal angle around the symmetry axis. We also define a curvilinear coordinate system comprising the $\mathrm{PEST}$ coordinates $(\psi_{\mathrm{p}}, \zeta, \theta)$ where $\psi_{\mathrm{p}}$ is the flux surface label, $\zeta$ is the cylindrical azimuthal angle and $\theta$ is the ``straight-field-line'' poloidal angle~\citep{d2012flux} such that $\alpha_{\mathrm{t}} = \zeta - q(\psi_{\rm{p}})( \theta - \theta_0)$. Similarly, for 3D equilibria, we use the coordinate system $(\psi, \zeta, \theta)$ and $\alpha_{\mathrm{s}} = \theta - \iota (\psi)(\zeta - \zeta_0)$. The pitch of the magnetic field line on a flux surface is described by the safety factor 
\begin{equation}
q(\psi) = \frac{1}{\iota(\psi)} \equiv  \frac{d \psi}{d \psi_{\mathrm{p}}} = \frac{1}{(2\pi)^2} \oint d\zeta \oint d\theta \, \frac{\boldsymbol{B}\bcdot \bnabla\zeta}{\boldsymbol{B}\bcdot \bnabla \theta},
\label{eqn:Safety-factor-definition}
\end{equation}
where $\iota$, the rotational transform, is the inverse of the safety factor. Using~\eqref{eqn:Div-free-B1} or~\eqref{eqn:Div-free-B2} for the magnetic field, one has to solve the steady-state ideal MHD force balance equation
\begin{equation}
     \boldsymbol{j} \times \boldsymbol{B}  = \boldsymbol{\nabla} p,
    \label{eqn:ideal MHD-force-balance}
\end{equation}
where $p$ is the plasma pressure, and $\boldsymbol{j}$ is the plasma current given by the Ampere's law
\begin{equation}
    \mu_0 \boldsymbol{j} = \bnabla \times \boldsymbol{B},
\end{equation}
where $\mu_0$ is the coefficient of permeability in vacuum. For axisymmetric equilibria, simplifying~\eqref{eqn:ideal MHD-force-balance} yields the Grad-Shafranov equation~\citep{grad1958proceedings, shafranov1957equilibrium}. For a general 3D equilibrium, such an equation does not exist. However, we can solve for the axisymmetric and 3D cases using a general numerical equilibrium solver. In the following section, we explain how we use the numerical solver~\texttt{VMEC}\footnote{The main idea of this work is independent of the equilibrium solver. Our technique should also work with any other equilibrium solver.} to obtain both 2D-axisymmetric and 3D equilibria.

\subsection{Numerical equilibrium solver}
\label{subsec:VMEC}
We generate numerical equilibria using the 3-D equilibrium code \texttt{VMEC}~\citep{hirshman1983steepest_VMEC}. The code works by minimizing the integral 
\begin{equation}
W = \int\left(\frac{p}{\gamma -1} + \frac{B^2}{2\,\mu_0}\right) dV,
\label{eqn:energy_integral}
\end{equation}
 subject to multiple geometric constraints~\citep{kruskal1958equilibrium}. For our study, we used the fixed-boundary mode of \texttt{VMEC}. The fixed-boundary mode takes the shape of the boundary surface denoted by the cylindrical coordinates $R_{\rm{b}}$ and $Z_{\rm{b}}$ in terms of the Fourier-decomposed poloidal ($\Theta$) and toroidal ($\zeta$) modes
\begin{equation}
\begin{gathered}
    R_{\rm{b}} =  \sum_{n}\sum_{m} \widehat{R}_{\rm{b}}(m, n) \exp(i(m \Theta - n \zeta)), \\
    Z_{\rm{b}} =  \sum_{n}\sum_{m} \widehat{Z}_{\rm{b}}(m, n) \exp(i(m \Theta - n \zeta)),
\end{gathered}
\label{eqn:Fourier-boundary}
\end{equation}
where $m$ and $n$ are integers. We also provide \texttt{VMEC} with the coefficients of the polynomials representing the global radial pressure $p(s)$ and the rotational transform $\iota(s)$ as a function of the normalized toroidal flux $s$, and the total toroidal or poloidal flux enclosed by the boundary. The poloidal angle $\Theta$ used by \texttt{VMEC} is related to the straight-field-line $\theta$ by the following equation 
\begin{equation}
    \Theta = \theta + \Lambda,
\end{equation}
where 
\begin{equation}
    \Lambda = \sum_{n}\sum_{m} \widehat{\Lambda}(m, n) \exp(i(m \Theta - n \zeta)).
\end{equation}
For a boundary shape, pressure, rotational transform, and enclosed toroidal flux, it then solves for the flux surfaces to minimize the integral in~\eqref{eqn:energy_integral} on each surface for fixed $p$ and $\iota$ subject to various topological constraints imposed by the ideal MHD.
In a more compact form, \texttt{VMEC} solves
\begin{align}
\min_{R, Z, \Lambda} W[R, Z, \Lambda; p, \iota, \psi(s=1)], \qquad \text{s.t. } R(s=1) = R_{\rm{b}}, \ Z(s=1) = Z_{\rm{b}}.
\end{align}
\iffalse
\begin{equation}
R^{*}(s), Z^{*}(s) = \argmin_{R,Z} W[R(s), Z(s), \Lambda(s); p, \iota, R_{\rm{b}}, Z_{\rm{b}}, \psi(s=1)],
\end{equation}
\fi
After running the code, we obtain the shape of the flux surfaces, the magnetic field, and a set of important physical quantities. The characteristic physical quantities that we will use in this work are defined below:
\begin{itemize}
    \item The total enclosed toroidal flux by the boundary $\psi_{\mathrm{b}} = 1/(2\pi) \int dV \, \boldsymbol{B}\cdot\bnabla \zeta$
    \item The normalizing magnetic field $B_{\mathrm{N}} = 2\, \psi_{\mathrm{b}}/(\pi a_{\mathrm{N}}^2)$ where $a_{\mathrm{N}} = \sqrt{\mathcal{A}_{\mathrm{b}}/\pi}$ is the effective minor radius and $\mathcal{A}_{\mathrm{b}}$ is the average area enclosed by the boundary
    \item The ratio of the total plasma pressure to the magnetic pressure on the magnetic axis $\beta_{\mathrm{ax}} = 2 \mu_0 p(s=0)/B_{\mathrm{N}}^2$
    \item The aspect ratio $A$ and the normalized minor radius $a_{\mathrm{N}}$ of the device 
    \item The radius of curvature of the boundary $R_c(\theta) = \frac{d^2\!R}{dZ^2}\big/(1 + (\frac{dR}{dZ})^2)^{3/2}$ where $R$ and $Z$ are the cylindrical coordinates used to parametrize the boundary
    \item The volume-averaged, normalized plasma pressure $\langle \beta \rangle = \int dV p/\int dV B^2$ where $dV$ is the differential volume element
    \item The total enclosed toroidal current $j_{\zeta}$ = $\lvert \int dV\, (\boldsymbol{j}\cdot \bnabla \zeta )\rvert$
    \item The volume-averaged magnetic field $\langle B \rangle = \int dV B/\sqrt{V}$ where $dV$ is the differential volume element
    \item The mean rotational transform $\bar{\iota} = \int ds \, \iota/\int ds$
\end{itemize}
Using \texttt{VMEC}, we generate three equilibria: an axisymmetric equilibrium with a D\textrm{III}-D-like boundary shape and two 3D equilibria: modified NCSX- and modified Henneberg-QA. In the following sections, we provide important details for each of these equilibria.

\subsection{2D axisymmetric equilibrium}
\label{subsec:2D-axisym-eqbm}
\begin{figure}
    \centering
    {\includegraphics[width=1.0\textwidth, trim = 0mm 0mm 0mm 0mm, clip]{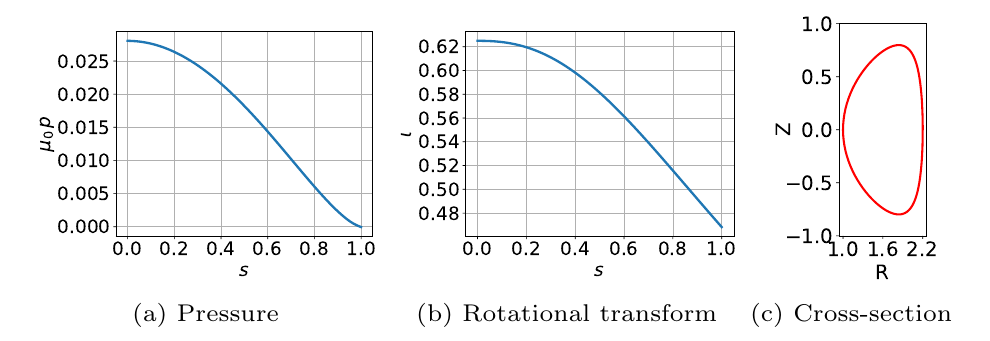}} 
    \caption{This figure plots the inputs to the \texttt{VMEC} code for the DIII-D-like design: the pressure, rotational transform as a function of the normalized toroidal flux $s$, and cross-section of the boundary.}
    \label{fig:DIII-D-input}
\end{figure}

In this study, the first equilibrium that we choose is a high-$\beta$, axisymmetric, D\textrm{III}-D-like equilibrium with a negative triangularity boundary -- a boundary that looks like an inverted-D. Negative triangularity equilibria have generally been found to have enhanced confinement by~\citet{marinoni2019h} while avoiding MHD-driven disruptions. We then choose a negative triangularity equilibrium from~\citet{gaur2022microstability} where it is shown to be unstable against the ideal ballooning mode. With this equilibrium as an initial state, we run our ideal ballooning stability optimization to find a stable equilibrium while maintaining a boundary shape with negative triangularity.
The input pressure, the rotational transform, and the boundary shape profile for this equilibrium are shown in figure~\ref{fig:DIII-D-input}.
Using these inputs, we run \texttt{VMEC} to obtain the global MHD equilibrium. 

Our Optimizer sometimes finds solutions that meet the ideal balloon stability constraints in a trivial fashion. For example, the optimizer may give us a large aspect ratio or decrease the minor radius causing the volume averaged $B$ to increase, causing the $\beta$ to decrease. To avoid these trivial solutions, we have to impose additional constraints on important characteristic physical quantities to prevent them from changing significantly. The values of the relevant physical quantities for this equilibrium are provided in table~\ref{tab:Table-1}.
\begin{table}
  \begin{center}
\def~{\hphantom{0}}
  \begin{tabular}{lcccccccc}
    $\beta_{\mathrm{ax}}(\%)$ &  $\langle \beta \rangle(\%)$  & $j_{\zeta} (\mathrm{M}A)$ & $\langle B \rangle$ (T)& $\bar{\iota}$ & $A$ & $a_{\mathrm{N}} (m)$ & $B_{\mathrm{N}}(T)$ & $\psi_{\mathrm{LCFS}}(T  m^2)$\\[3pt]
    $14$  & $7.6$  & $0.616 $ & $0.677$ & $0.568$ & $2.42$ & $0.68$ & $0.679$ & $1.0$ \\
  \end{tabular}
    \caption{This table shows the values of relevant physical quantities for the DIII-D like equilibrium.}
  \label{tab:Table-1}
  \end{center}
\end{table}

\subsection{Modified NCSX equilibrium}
\label{subsec:3D-NCSX-eqbm}
\begin{figure}
    \centering
    {\includegraphics[width=1.0\textwidth, trim = 0mm 0mm 0mm 0mm, clip]{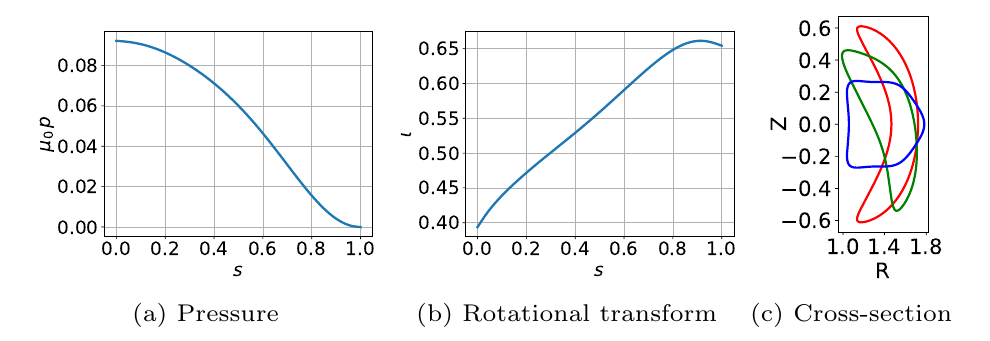}} 
    \caption{This figure plots the inputs to the \texttt{VMEC} code for the modifief NCSX design: the pressure, rotational transform as a function of the normalized toroidal flux $s$, and cross-section of the boundary. Notice the large negative shear until $s \approx 0.85$.}
  \label{fig:NCSX-input}
\end{figure}

The second equilibrium we select is the 3D equilibrium for the NCSX design~\citep{fu2007ideal, zarnstorff2001physics}. This equilibrium is designed to have a hidden symmetry known as quasisymmetry~\citep{garren1991magnetic, boozer1983transport} where the strengh of the magnetic field $|\boldsymbol{B}|$ does not change along the field line with respect to one of the coordinates. Quasisymmetry is a useful property because it ensures orbit confinement, which helps improve energetic particle confinement, a major issue in stellarators. 
The pressure, rotational transform, and boundary shape profile for this equilibrium are shown in figure~\ref{fig:NCSX-input}.

In tokamaks, magnetic field lines are twisted through an externally induced toroidal plasma current $j_{\zeta}$. On the other hand, in stellarators, the fieldlines are twisted using a non-axisymmetric boundary shape instead of inducing a plasma current. The 3D boundary shape alleviates the need for an internal current and eliminates all current-driven instabilities known to cause disruptions in tokamaks. However, because of neoclassical effects, stellarator equilibria can still generate some toroidal current. To eliminate the need for an externally induced current, one must ensure that the bootstrap current is self-consistent with the 3D equilibrium. Since we are not optimizing for a self-consistent bootstrap current, it is essential to ensure that the toroidal current does not have an enormous value. Therefore, we include additional constraints to prevent large changes in the toroidal current and other important quantities. The values of these equilibrium-dependent quantities are presented in table~\ref{tab:Table-2}.

\begin{table}
  \begin{center}
\def~{\hphantom{0}}
  \begin{tabular}{lcccccccc}
    $\beta_{\mathrm{ax}}(\%)$ &  $\langle \beta \rangle(\%)$  & $j_{\zeta} (\mathrm{M}A)$ & $\langle B \rangle$ (T)& $\bar{\iota}$ & $A$ & $a_{\mathrm{N}} (m)$ & $B_{\mathrm{N}}(T)$ &$\psi_{\mathrm{LCFS}}(T  m^2)$\\[3pt]
    $9.38$  & $5.14$  & $0.174$ & $1.59$ & $0.55$ & $4.36$ & $0.32$ & $1.54$ & $0.514$ \\
  \end{tabular}
    \caption{This table shows the values of important physical quantities for the modified NCSX equilibrium.}
  \label{tab:Table-2}
  \end{center}
\end{table}

\subsection{Modified Henneberg-QA}
\label{subsec:3D-HBERG-eqbm}
\begin{figure}
    \centering
    {\includegraphics[width=1.0\textwidth, trim = 0mm 0mm 0mm 0mm, clip]{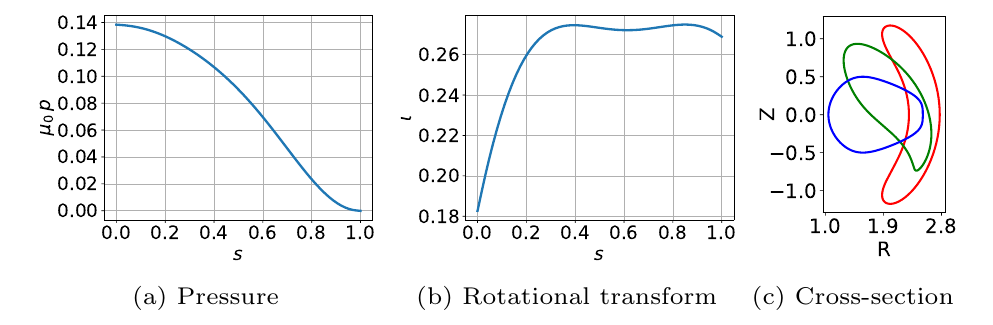}} 
    \caption{This figure plots the inputs to the \texttt{VMEC} code for the modified Henneberg-QA design: the pressure, rotational transform as a function of the normalized toroidal flux $s$, and cross-section of the boundary. Notice the large negative shear in the inner-core region.}
    \label{fig:HBERG-input}
\end{figure}

The final equilibrium we study is the modified Henneberg-QA design~\citep{HennebergQA}. This equilibrium is also designed to have quasisymmetry for a wide variety of pressure profiles. The pressure, rotaional transform and boundary shape profile for this equilibrium are shown in figure~\ref{fig:HBERG-input}. 

For reasons explained in the previous section, we will impose additional constraints on some of the physical quantities. The values of these equilibrium-dependent parameters that we will use as constraints in~\S\ref{sec:results} are presented in table~\ref{tab:Table-3}.
\begin{table}
  \begin{center}
\def~{\hphantom{0}}
  \begin{tabular}{lcccccccc}
    $\beta_{\mathrm{ax}}(\%)$ &  $\langle \beta \rangle (\%)$  & $j_{\zeta} (\mathrm{M}A)$ & $\langle B \rangle (T)$& $\bar{\iota}$ & $A$ & $a_{\mathrm{N}} (m)$ & $B_{\mathrm{N}}\!(T)$  &$\psi_{\mathrm{LCFS}}(T  m^2)$\\[3pt]
    $4.5$  & $2.4$  & $0.235$ & $2.5$ & $0.263$ & $3.37$ & $0.60$ & $2.35$ & $2.67$ \\
  \end{tabular}
    \caption{This table presents values of relevant physical quantities for the modified Henneberg-QA design.}
  \label{tab:Table-3}
  \end{center}
\end{table}

In the following section, we describe the ideal ballooning stability and analyze these equilibria by measuring their instability against the ideal ballooning mode.

\section{The infinite-$n$ ideal ballooning mode}\label{sec:ideal-balloning mode}
In this section, we present the details of the infinite-$n$ ideal ballooning mode. In~\S\ref{subsec:Phy-and-math-description}, we briefly describe its theoretical foundation and mathematical formulation. In~\S\ref{subsec:ball-growth-rates}, we present a numerical technique used to solve the ideal ballooning equation. In~\S\ref{subsec:iball-self-adjoint}, we describe the mathematical properties that we use to formulate an adjoint-based method and accelerate optimization against the ideal ballooning mode. In the final section, we will describe how optimization against the ideal ballooning mode can speed up optimization against an electromagnetic mode seen in kinetic plasma turbulence, known as the Kinetic Ballooning Mode (KBM).
\subsection{Physical and mathematical description}
\label{subsec:Phy-and-math-description}
This work involves a detailed analysis of three equilibria against an important MHD instability, the infinite-$n$ ideal ballooning instability \citep{ConnorHastieTaylorProcRoySoc, DewarGlasserballooning} --- a field-aligned, pressure-driven Alfv\'{e}n wave that grows when the destabilizing pressure gradient in the region of ``bad'' curvature exceeds the stabilizing effect of field-line bending. Using magnetic field unit vector $\boldsymbol{b} = \boldsymbol{B}/B$, the region of ``bad'' curvature is defined as a region of a flux surface where $(\boldsymbol{b}\cdot\bnabla \boldsymbol{b}) \cdot \bnabla p > 0$. 

The ideal ballooning equation  
\begin{equation}
    \frac{1}{\mathcal{J}}\frac{\partial}{\partial \theta}\left( \frac{\lvert \bnabla\alpha_{\rm{t}} \rvert^2 }{\mathcal{J}\, B^2} \frac{\partial \widehat{X}}{\partial \theta}\right) + 2 \frac{dp}{d\psi}\left[ \boldsymbol{B}\times\bnabla\left(\mu_0 p + \frac{B^2}{2}\right)\bcdot\bnabla{\alpha_{\rm{t}}}\right] \widehat{X} = -\rho \omega^2 \frac{\lvert\bnabla\alpha_{\rm{t}}\rvert^2}{B^2} \widehat{X}, 
    \label{eqn:ballooning-equation}
\end{equation}
is a second-order eigenvalue differential equation that calculates the perturbation $\widehat{X}(\theta)$ along the ballooning coordinate $\theta$ and its eigenvalue (or growth rate) $-\omega^2$. In~\eqref{eqn:ballooning-equation}, $\rho$ is the plasma mass density, $\mathcal{J} = (\boldsymbol{B}\cdot \bnabla \theta)^{-1}$ and the rest of the terms are defined in~\S\ref{sec:equilibrium}. This equation is solved subject to the following condition on the eigenfunction
\begin{equation}
    \lim_{\theta \rightarrow \pm \infty} \widehat{X}(\theta; \psi, \alpha_{\rm{t}}, \theta_0) = 0,
\end{equation}
where $\theta_0$ is the ballooning parameter\footnote{In the context of infinite-$n$ ideal ballooning mode analyses, there is a value of the ballooning parameter $\theta_0$ at which the ballooning mode is the least stable. To find this value, we treat $\theta_0$ as a free parameter and scan its values to find $\theta_0$ for which $\omega^2$ is the most negative}. Details of the derivation of the ideal ballooning equation are given in~\cite{ConnorHastieTaylorProcRoySoc, DewarGlasserballooning}.

The ballooning equation balances the stabilizing fieldline bending term and the destabilizing pressure gradient with the inertia of the resulting Alfv\'{e}n wave, oscillating with a frequency $\omega$. Note that~\eqref{eqn:ballooning-equation} depends on $\psi$ (or $\psi_{\rm{p}}$) as a parameter, and we can compute the coefficients from the equilibrium quantities on each surface. Before solving~\eqref{eqn:ballooning-equation} numerically, we normalize and write the ballooning equation on a fieldline (fixed $\alpha_{\rm{t}}$) as
\begin{equation}
    \frac{d}{d\theta}\mathrm{g} \frac{d \widehat{X}}{d\theta} + \mathrm{c} \widehat{X} = \widehat{\lambda} \mathrm{f} \widehat{X},
    \label{eqn:iball-condensed}
\end{equation}
where 
\begin{equation}
\begin{gathered}
    \mathrm{g} = (\boldsymbol{b}\cdot\bnabla_{\mathrm{N}} \theta) \frac{\lvert \bnabla_{\mathrm{N}}\alpha_{\rm{t}} \rvert^2}{B/B_{\rm{N}}},\\
    \mathrm{c} = \frac{2}{(\boldsymbol{B}\cdot\bnabla_{\mathrm{N}} \theta)}\, \frac{d (\mu_0 p/B_{\rm{N}}^2)}{d\psi_{\mathrm{N}}}\left[\boldsymbol{B} \times \bnabla_{\mathrm{N}} \left(\frac{2\mu_0 p + B^2}{2 B_{\rm{N}}^2}\right) \cdot \bnabla_{\rm{N}} \alpha_{\rm{t}} \right],\\
    \mathrm{f} =  \frac{\lvert \bnabla_{\mathrm{N}} \alpha_{\rm{t}} \rvert ^2}{(B/B_{\rm{N}})^2},\\
    \widehat{\lambda} = -\left(\frac{\omega a_{\mathrm{N}}}{v_{\mathrm{A}}}\right)^2,\quad v_{\mathrm{A}} = \frac{B_{\mathrm{N}}}{\sqrt{4\pi \rho}}, 
\end{gathered}
\end{equation}
where $v_{\mathrm{A}}$ is the Alfv\'{e}n speed and the values and definitions of the effective minor radius $a_{\mathrm{N}}$ and the normalizing magnetic field $B_{\mathrm{N}}$ are the normalizing length and mangetic field strength, respectively, defined in~\S\ref{subsec:VMEC}. The ideal ballooning equation is solved subject to the boundary conditions
\begin{equation}
 \widehat{X}(\theta=\pm \theta_{\mathrm{b}}; \psi, \alpha_{\rm{t}}, \theta_0)  = 0.   
\end{equation}
where $\theta_{\mathrm{b}}$ is a finite user-selected value that determines the extent of the eigenfunction. In the next section, we present the numerical procedure used to solve the ideal ballooning equation.

\subsection{Numerical implementation and eigenvalues of the selected equilibria} 
\label{subsec:ball-growth-rates}
In this section, we will briefly discuss the numerical technique used to solve the ballooning equation~\eqref{eqn:iball-condensed}. Our numerical technique is virtually identical to that used by~\citet{sanchez2000cobra} in their ballooning solver \texttt{COBRAVMEC}. After briefly explaining the details of our solver, we present the maximum eigenvalue as a function of the normalized toroidal flux $s$ for the three equilibria we presented in~\S\ref{sec:equilibrium}. 

The ideal ballooning equation is a second-order ordinary differential equation with real-valued coefficients. This eigenvalue equation is discretized using a second-order accurate, central-finite-difference scheme
\begin{equation}
    \mathrm{g}_{j+1/2}\frac{(\widehat{X}_{j+1} - \widehat{X}_{j})}{ \Delta \theta^2} - \mathrm{g}_{j-1/2}\frac{(\widehat{X}_{j} - \widehat{X}_{j-1})}{\Delta \theta^2} + (\mathrm{c}_j - \widehat{\lambda} \mathrm{f}_j)\widehat{X}_j = 0, \quad j = 0\ldots N-1
    \label{eqn:discrete-iball}
\end{equation}
where $N$ is an odd number of uniformly spaced points in the ballooning space, $\theta_j \in [-\theta_{\mathrm{b}}, \theta_{\mathrm{b}}]$ and $\Delta \theta = \theta_{j+1} - \theta_{j}$. First-order derivatives are evaluated at half points $j-1/2, j-3/2$ instead of grid points to ensure numerical stability. The boundary conditions satisfied by the discrete equations are $\widehat{X}_0 = \widehat{X}_{N} = 0$. 
For a fixed poloidal and toroidal resolution, the time taken by our solver is proportional to $\theta_{\mathrm{b}}$. Therefore, it is important to find the right balance between speed and accuracy. Throughout this work, we chose the domain limit $\theta_{\mathrm{b}} = 5\pi$ for all calculations, as we found it to be a sufficient range to capture the ballooning eigenfunction. We observed that the values $\theta_{\mathrm{b}} > 5\pi$ made a relatively small difference from the value obtained of $\widehat{\lambda}$. The discrete set of equations~\eqref{eqn:discrete-iball} is written in the form of a matrix equation
\begin{equation}
    A \widehat{X} = \widehat{\lambda} \widehat{X},
    \label{eqn:matrix-iball-eqn}
\end{equation}
where the exact matrix $A$ is provided in appendix~\ref{app:ideal_ballooning_matrix}. We then solve~\eqref{eqn:matrix-iball-eqn} to find the largest eigenvalue using an Arnoldi iterative scheme using the \texttt{scipy.linalg.eigs} solver in \texttt{Python} and refine the accuracy of the largest eigenvalue in the grid spacing $\Delta \theta$ using variational refinement 
\begin{equation}
        \widehat{\lambda} = \frac{\bigintss_{-\theta_{\rm{b}}}^{\theta_{\rm{b}}} d\theta \left( \mathrm{c} \lvert \widehat{X}\rvert^2 -  \mathrm{g} \Big\lvert \dfrac{d\!\widehat{X}}{d\theta}\Big\rvert^2 \right)}{\bigintssss_{-\theta_{\rm{b}}}^{\theta_{\rm{b}}} d\theta\,  \mathrm{f}\lvert \widehat{X}\rvert^2},
\end{equation}
where the derivative $d\!\widehat{X}/d\theta$ is calculated using a fourth-order accurate finite difference scheme and the integral is performed using a fourth-order accurate Simpson's rule ($1/3$ rule) with \texttt{scipy.integrate.simps}\footnote{Our \textrm{Python} code is freely available at \href{https://github.com/rahulgaur104/ideal-ballooning-solver}{github.com/rahulgaur104/ideal ballooning-solver}}. Note that we only solve for and refine the largest eigenvalue of~\eqref{eqn:iball-condensed} and not the entire eigenvalue spectrum.

\subsection{Properties of the ideal ballooning equation}
\label{subsec:iball-self-adjoint}
The ideal ballooning~\eqref{eqn:iball-condensed}  is a linear equation that can be written as 
\begin{equation}
    \mathcal{L}\widehat{X} = \widehat{\lambda} \widehat{X},
    \label{eqn:gen-eigenvalue-eqn}
\end{equation}
where the linear operator 
\begin{equation}
    \mathcal{L} \equiv \frac{1}{\mathrm{f}}\frac{d}{d\theta}\mathrm{g}\frac{d}{d\theta} + \frac{\mathrm{c}}{\mathrm{f}}, 
    \label{eqn:iball-condensed-2}
\end{equation}
and the coefficients $\mathrm{g, c, f}$ are real-valued functions along a field line. Mathematically, the solutions of~\eqref{eqn:gen-eigenvalue-eqn} form the basis for the Hilbert space equipped with the following inner product
\begin{equation}
    \langle \widehat{X}_1, \widehat{X}_2 \rangle = \int_{-\infty}^{\infty} d\theta \widehat{X}_1^{*} \widehat{X}_2  , 
\end{equation}
and are square integrable, i.e., $\langle \widehat{X}, \widehat{X} \rangle \! <\! \infty$. Due to the self-adjoint nature of ideal MHD~\citep{IdealMHD}, for solutions $\widehat{X}_1$ and $\widehat{X}_2$ of equation~\eqref{eqn:gen-eigenvalue-eqn} the operator $\mathcal{L}$ satisfies the following property
\begin{equation}
    \langle \mathcal{L} \widehat{X}_1, \widehat{X}_2 \rangle  = \langle \widehat{X}_1, \mathcal{L}\widehat{X}_2 \rangle,
    \label{eqn:self-adjoint-property1}
\end{equation}
where we have used the boundary condition $\lim_{\theta \rightarrow \pm \infty} \widehat{X}_1 = \lim_{\theta \rightarrow \pm \infty} \widehat{X}_2 = 0$. Using~\eqref{eqn:self-adjoint-property1}, one can show that all eigenvalues $\widehat{\lambda}$~\eqref{eqn:ballooning-equation} will be real numbers. Therefore, $\omega = \pm i\sqrt{\widehat{\lambda}}$ will be purely real, an oscillating mode, or purely imaginary, a growing mode. We refer to oscillating modes as stable and to growing modes as unstable. We will use this powerful property in~\S\ref{sec:Adjoint-method} to formulate an adjoint method, a technique that can speed up the calculation of the gradient of $\widehat{\lambda}_{\mathrm{max}}$ on each flux surface.

\subsection{Relation to the Kinetic Ballooning Mode}
In this section, we explain how the ideal ballooning equation is directly related to an important mode of gyrokinetic plasma turbulence known as the Kinetic Ballooning Mode (KBM). Unlike ideal MHD which is a fluid theory, a single fluid with properties that vary in configuration space, a gyrokinetic model takes into account the distribution of different ion and electron species in both configuration and velocity space. Using a gyrokinetic model,~\citet{tang1980kinetic} have shown that for devices with a large aspect ratio, modes with wavenumbers $k_y 
 \ll  1/\rho_{\mathrm{i}}$, where $\rho_i = e B/m_{\mathrm{i}}$ is the ion gyroradius and $m_{\mathrm{i}}$ is the ion-mass, the gyrokinetic model can be reduced to the ideal ballooning equation with corrections that depend on $k_y\rho_{\mathrm{i}}$.
\begin{equation}
    \frac{d}{d\theta}\mathrm{g} \frac{d \widehat{X}}{d\theta} + \mathrm{c} \widehat{X} = \omega (\omega_{*,s} - \omega) \mathrm{f} \widehat{X},
\end{equation}
where
\begin{equation}
    \omega_{*,s} = c_{0} \, k_y \rho_{\mathrm{i}},
\end{equation}
where $c_0$ is a constant on a flux surface and $k_y \rho_{\mathrm{i}}$ is the normalized wavenumber of the mode. 
\iffalse
A brief self-contained calculation is required to derive equation \textcolor{red}{number the equation} is given in the appendix \textcolor{red}{Appendix-name}. 
\fi
In the long wavelength limit, i.e., $k_y \rho_i \rightarrow 0$, we recover the ideal ballooning equation exactly. This means that an ideal ballooning unstable mode is also unstable to the KBM. In fact, the KBM is an ideal ballooning mode with kinetic effects. 

Using a simple mixing-length argument, one can qualitatively argue that the turbulence heat flux diffusion is
\begin{equation}
    D \sim \frac{\mathrm{Im}(\omega)}{k_y^2},
\end{equation}
where $\mathrm{Im}(\omega)$ is the imaginary part of $\omega$ also known as the growth rate. This implies that low-wavenumber turbulence has the highest rate of diffusion and leads to poor plasma confinement. Hence, even if ideal ballooning unstable modes do not lead to disruption in stellarators, they could lead to a large heat flux transport through the KBM channel. 
Since calculating KBM growth rates using a microstability code is expensive, one can use ideal ballooning stability as a necessary condition for KBM stability to optimize against the KBM~\footnote{Note that ideal ballooning stability is a necessary but not sufficient condition for KBM stability. A mode can be stable against the ideal ballooning mode, but unstable against the KBM}. For tokamaks, this is one of the fundamental ideas currently used in the~\texttt{EPED} code~\citep{snyder2007stability} to predict the plasma pressure profile in the pedestal region.

In summary, in this section, we have explained the mathematical, physical, and numerical methods used to solve the ideal ballooning equation. We have also explained the self-adjoint property of the ideal ballooning equation and the crucial link between the ideal and kinetic ballooning modes. In the next section, we will use the self-adjoint property of the ideal ballooning equation to outline and test the adjoint method.

\section{Developing an adjoint method}\label{sec:Adjoint-method}
In this section, we derive and explain the process of calculating the gradients of the ideal ballooning eigenvalue $\widehat{\lambda}$ on each surface that would help us find the maximum eigenvalue $\widehat{\lambda}_{\rm{max}}$ using an adjoint method. We then elucidate how it is faster than the conventional gradient-based method and illustrate this by plotting gradients from a typical optimization run and calculating the speed-up.

To find $\widehat{\lambda}_{\rm{max}}$ on each flux surface, we need the gradient of the eigenvalue of a general function $\mathcal{H}$ such that $\mathcal{H}$ is maximized if and only if $\widehat{\lambda} = \widehat{\lambda}_{\rm{max}}$. Mathematically, this problem can be defined as follows
\begin{equation}
    \max \mathcal{H}(\widehat{\lambda}, \widetilde{\boldsymbol{p}}, \widehat{\boldsymbol{p}}), \qquad \textrm{s.t.} \quad \mathcal{G}(\widehat{\lambda}, \widehat{X}, \widetilde{\boldsymbol{p}}, \widehat{\boldsymbol{p}}) \equiv \mathcal{L} \widehat{X} - \widehat{\lambda} \widehat{X} =  0, 
\end{equation}
where $\widehat{\lambda}$ is the eigenvalue, $\widehat{X}$ is the eigenfunction, $\widetilde{\boldsymbol{p}}$ is the state vector that contains all the equilibrium parameters such as the boundary shape and the $\iota$ profile and $\widehat{\boldsymbol{p}} = (\alpha_{\rm{t}}, \theta_0)$ is a vector that contains the parameters of the ideal ballooning equation, $\mathcal{H}$ is an objective function, and $\mathcal{G}$ is the ideal ballooning operator. To maximize $\mathcal{H}$ on a flux surface for a given equilibrium, i.e., for a fixed $\widetilde{\boldsymbol{p}}$, we need the gradient
\begin{equation}
    \frac{d \mathcal{H}}{d \widehat{\boldsymbol{p}}} = \frac{\partial \mathcal{H}}{\partial \widehat{\lambda}}\bigg\lvert_{\widehat{\boldsymbol{p}}} \frac{\partial \widehat{\lambda}}{\partial \widehat{\boldsymbol{p}}} + \frac{\partial \mathcal{H}}{\partial \widehat{\boldsymbol{p}}}\bigg\lvert_{\widehat{\lambda}}.
    \label{eqn:dhdp}
\end{equation}
The most expensive term to calculate in~\eqref{eqn:dhdp} is the gradient of the eigenvalue $\lambda$. To obtain that, we take the derivative of the operator $\mathcal{G}$ with respect to $\lambda$\footnote{Note that the derivative of an eigenvalue is only well-defined when the eigenvlaue is isolated. Optimization problems with stringent penalty terms can lead the optimizer to points with multiplicity~\citep{lewis1996eigenvalue}.} for a fixed $\widetilde{\boldsymbol{p}}$
\begin{equation}
    -\frac{\partial \mathcal{G}}{\partial \widehat{\lambda}}\frac{\partial \widehat{\lambda}}{\partial \widehat{\boldsymbol{p}}} = \frac{\partial \mathcal{G}}{\partial \widehat{X}} \frac{\partial \widehat{X}}{\partial \widehat{\boldsymbol{p}}} + \frac{\partial \mathcal{G}}{\partial \widehat{\boldsymbol{p}}}\bigg\lvert_{\widehat{X}, \widehat{\lambda}}.
    \label{eqn:dGdp}
\end{equation}
This equation can be explicitly written with the help of~\eqref{eqn:iball-condensed-2}
\begin{equation}
    \frac{\partial \widehat{\lambda}}{\partial \widehat{\boldsymbol{p}}} \widehat{X} = (\mathcal{L}-\widehat{\lambda}) \frac{\partial \widehat{X}}{\partial \widehat{\boldsymbol{p}}} + \frac{\partial \mathcal{L}}{\partial \widehat{\boldsymbol{p}}}\widehat{X}.
    \label{eqn:dGdp1}
\end{equation}
To simplify~\eqref{eqn:dGdp1} further, we multiply it by the eigenfunction $\widehat{X}^{*}$ and integrate it throughout the domain $\theta \in [-\theta_{\rm{b}}, \theta_{\rm{b}}]$. Upon doing that, we use the self-adjoint property~\eqref{eqn:self-adjoint-property1} and work through the algebra (given in appendix~\ref{app:adjoint-iball-math}) to obtain the adjoint relation
\begin{equation}
        \frac{\partial \widehat{\lambda}}{\partial \widehat{\boldsymbol{p}}} = \frac{\bigintss_{-\theta_{\rm{b}}}^{\theta_{\rm{b}}} d\theta \left(\dfrac{\partial \mathrm{c}}{\partial \widehat{\boldsymbol{p}}} \lvert \widehat{X}\rvert^2 - \dfrac{\partial \rm{g}}{\partial \widehat{\boldsymbol{p}}}\Big\lvert \dfrac{d\widehat{X}}{d\theta}\Big\rvert^2 -  \widehat{\lambda} \dfrac{\partial \mathrm{f}}{\partial \widehat{\boldsymbol{p}}} \lvert \widehat{X} \rvert^2 \right)}{\bigintssss_{-\theta_{\rm{b}}}^{\theta_{\rm{b}}} d\theta\,  \mathrm{f}\lvert \widehat{X}\rvert^2}.
        \label{eqn:dlamdp-main}
\end{equation}
To obtain $\partial \widehat{\lambda}/\partial \widehat{\boldsymbol{p}}$ using a central finite difference scheme, one has to solve the ideal ballooning equation $2 n_{\widehat{\boldsymbol{p}}} = 4$ times at each optimization step, where $n_{\widehat{\boldsymbol{p}}}$ is the length of the vector $\widehat{\boldsymbol{p}}$. However, using the adjoint relation~\eqref{eqn:dlamdp-main}, we only have to solve it once per optimization step, as long as we can calculate the gradients of geometry-related quantities $ \mathrm{g}, \mathrm{c}$, and $\mathrm{f}$ four times. Since gradients of $\rm{g}, \rm{c}, \rm{f}$ can be calculated roughly two orders of magnitude faster than solving the ideal ballooning equation, we speed up the gradient calculation by approximately a factor of four. Therefore, we use the adjoint relation~\eqref{eqn:dlamdp-main} to calculate the gradient of $\widehat{\lambda}$. In this study, we choose
\begin{equation}
    \mathcal{H}(\widehat{\lambda}, \widetilde{\boldsymbol{p}}, \widehat{\boldsymbol{p}}) = \widehat{\lambda}.
\end{equation}
Applying this fact to~\eqref{eqn:dhdp} and using~\eqref{eqn:dlamdp-main},
\begin{equation}
    \frac{d \mathcal{H}}{d\widehat{\boldsymbol{p}}} = \frac{\partial \widehat{\lambda}}{\partial \widehat{\boldsymbol{p}}} = \frac{\bigintss_{-\theta_{\rm{b}}}^{\theta_{\rm{b}}} d\theta \left(\dfrac{\partial \mathrm{c}}{\partial \widehat{\boldsymbol{p}}} \lvert \widehat{X}\rvert^2 - \dfrac{\partial \rm{g}}{\partial \widehat{\boldsymbol{p}}}\Big\lvert \dfrac{d\widehat{X}}{d\theta}\Big\rvert^2 -  \widehat{\lambda} \dfrac{\partial \mathrm{f}}{\partial \widehat{\boldsymbol{p}}} \lvert \widehat{X} \rvert^2 \right)}{\bigintssss_{-\theta_{\rm{b}}}^{\theta_{\rm{b}}} d\theta\,  \mathrm{f}\lvert \widehat{X}\rvert^2}.
    \label{eqn:dhdp2}
\end{equation}
This relation gives us the derivative of the ballooning objective function at any point $\widehat{\boldsymbol{p}} = (\alpha_{\rm{t}}, \theta_0)$. Note that in this work we will use~\eqref{eqn:dhdp2} to find $\widehat{\lambda}_{\mathrm{max}}$ on a flux surface. However, this method is valid and, under appropriate conditions, can be extended to the equilibrium parameters $\widetilde{\boldsymbol{p}}$. The details of the extended adjoint method are given in appendix~\ref{app:adjoint-iball-general}.

In the next section, we present data and explain the advantages of adjoint methods over the regular finite-difference-based method to calculate the gradients of $\widehat{\lambda}_{\mathrm{max}}$.

\subsection{Comparing adjoint gradients with a finite difference method}
\label{subsec:adj-comparison}
\begin{figure}
    \centering
    {\includegraphics[width=0.85\textwidth, trim = 0mm 1mm 0mm 0mm, clip]{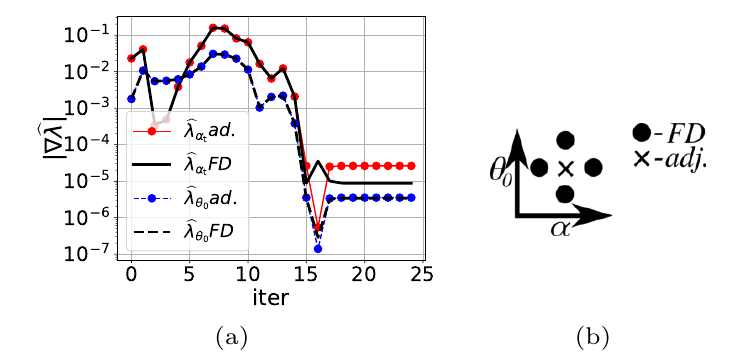}} 
    \caption{In this figure we present ($\mathit{a}$) comparison between the gradients of eigenvalue $\widehat{\lambda}_{\alpha_{\mathrm{t}}} = \partial \widehat{\lambda}/\partial \alpha_{\mathrm{t}}$ and $\widehat{\lambda}_{\theta_0} = \partial \widehat{\lambda}/\partial \theta_0$ obtained using a finite difference scheme against ones obtained using an adjoint method. The quantity $\mathrm{iter}$ is the number of iterations taken by the local optimizer on a flux surface before finding $\widehat{\lambda}_{\mathrm{max}}$. The gradients match well for around four orders of magnitude. The discrepancy between the adjoint and finite difference $\widehat{\lambda}_{\alpha_{\mathrm{t}}}$ is due to the finite resolution of the $\texttt{VMEC}$ run. In figure ($\mathit{b}$), we illustrate the the different grids used to calculate the gradient of the eigenvalue $\widehat{\lambda}$ on a flux surface. A finite difference scheme requires four points whereas an adjoint method only requires one points. This gives us a four times speed-up.}
    \label{fig:adjoint-FD-comparison}
\end{figure}

In this section, we will first compare the values of the gradients of $\lambda_{\mathrm{max}}$ from the adjoint method with their values obtained using a central-finite difference method. We take a typical optimization loop in the modified NCSX case and show a gradient comparison in figure~\ref{fig:adjoint-FD-comparison}. As you can see, the gradients obtained using an adjoint method match well with the gradients obtained with a finite-difference method.

To show the computational speedup, we also compare the time taken by an adjoint method with the regular finite-difference-based method. For the thirty iterations shown in figure~\ref{fig:adjoint-FD-comparison}($\mathit{a}$), the adjoint method was about $4$ times faster than the finite difference method. Indeed, the most expensive part of the gradient calculation is the ballooning solver. As shown in the illustration in figure~\ref{fig:adjoint-FD-comparison}($\mathit{b}$), for a second-order accurate central-difference scheme, an adjoint method only needs a single call to the ballooning solver, whereas the finite difference solver needs four. In principle, a speed-up factor of up to $4$ should be possible.

\section{Details of the optimization process}
\label{sec:optimizn-details}
In this section, we will explain the optimization process to find equilibria that are stable against the ideal ballooning mode. In~\S\ref{subsec:Max-lambda-optimization}, we describe the process of using an adjoint method to find the maximum growth rate $\widehat{\lambda}_{\mathrm{max}}$ on each flux surface. In~\S\ref{subsec:ball-optimization}, we then explain how we use $\widehat{\lambda}_{\mathrm{max}}$ and other penalty terms to construct the overall objective function $\mathcal{F}$. Finally, in~\S\ref{subsec:SIMSOPT}, we explain how we search for ballooning stable equilibria while minimizing $\mathcal{F}$ using the \texttt{SIMSOPT} framework. 

\subsection{Finding $\widehat{\lambda}_{\mathrm{max}}$ on each flux surface}
\label{subsec:Max-lambda-optimization}
\begin{figure}
    \centering
    {\includegraphics[width=0.85\textwidth, trim = 0mm 0mm 0mm 0mm, clip]{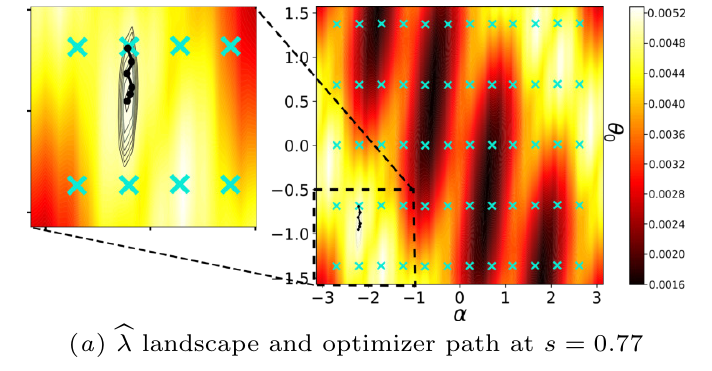}} 
    \caption{This figure shows the typical process of finding the globally maximum eigenvalue $\widehat{\lambda}_{\mathrm{max}}$ on the flux surface $s = 0.77$. We start by first finding the maximum $\widehat{\lambda}$ on discrete grid of $\alpha_{\rm{t}}$ and $\theta_0$. From the maximum discrete $\widehat{\lambda}$, we search for the global maximum eigenvalue using a local optimizer. In the inset, we show the approximate path taken by the optimizer to reach the $\widehat{\lambda}_{\mathrm{max}}$. }
    \label{fig:max-lambda-search-illustration}
\end{figure}

\begin{figure}
    \centering
    {\includegraphics[width=1.0\textwidth, trim = 2mm 1mm 4mm 0mm, clip]{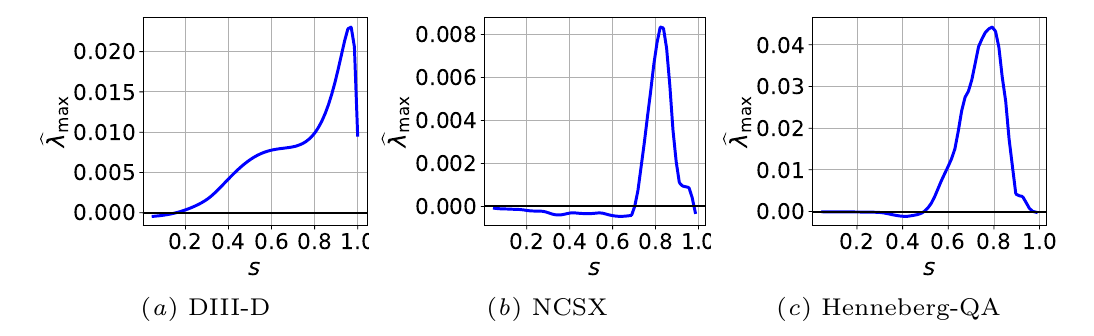}} 
    \caption{This figure shows the plots of $\widehat{\lambda}_{\mathrm{max}}$ against the normalized toroidal flux $s$ for the three chosen equilibria. Note that the ideal ballooning growth rate $\omega = i \sqrt{\widehat{\lambda}}$ so $\widehat{\lambda}_{\mathrm{max}} = 0.008$ corresponds to $\omega = i 0.09$ which is not a small growth rate.}
    \label{fig:max-lambda-plots}
\end{figure}

To calculate the ballooning objective function we find the maximum $\widehat{\lambda}$ on each flux surface. To do that, we solve~\eqref{eqn:iball-condensed} on several flux surfaces, multiple field lines on each surface, and numerous values of $\theta_0$ on each field line. 
We calculate $\widehat{\lambda}_{\mathrm{max}}$ on $\mathrm{ns} = 16$ flux surfaces for each equilibrium. For the 3D equilibria, we scan $n_{\alpha_{\rm{t}}} = 42$ field lines in the range $\alpha_{\rm{t}} = [-\pi, \pi)$. Since all field lines are identical in a 2D axisymmetric equilibrium, we scan only one field line, i.e., $n_{\alpha} = 1$ for the 2D equilibrium. On each field line, we scan $n_{\theta_0} = 21$ values of $\theta_0$ in the range $\theta_0 = [-\pi/2, \pi/2)$. The maximum $\widehat{\lambda}$ from a coarse grid scan gives us a value close to the global maximum. From the maximum $\widehat{\lambda}$ of the coarse grid, we launch a local gradient-based optimizer to find the global maximum eigenvalue. This process is explained using the illustration~\ref{fig:max-lambda-search-illustration}.
Using this process, we obtain $\widehat{\lambda}_{\mathrm{max}}$ as a function of the normalized toroidal flux $s$. Figure~\ref{fig:max-lambda-plots} shows the plot of $\widehat{\lambda}_{\mathrm{max}}$ against $s$ for the three chosen equilibria.

For each new equilibrium, on all $\mathrm{ns}=16$ flux surfaces, the local optimizer takes an average of $20$ iterations to find $\widehat{\lambda}_{\mathrm{max}}$. Moreover, as described in figure~\ref{fig:adjoint-FD-comparison}$\mathit(b)$, at each step, the use of a finite difference method requires $4$ evaluations of the eigenvalue $\widehat{\lambda}$. This means that on average, we have to call the ballooning solver $1280$ ($16 \times 20 \times 4$) times. This is a computationally expensive step that we speed up using our adjoint-based method.

\subsection{Finding ballooning-stable equilibria}
\label{subsec:ball-optimization}
Once we have found $\widehat{\lambda}_{\mathrm{max}}$, we seek an equilibrium stable to the ideal ballooning mode by minimizing $\widehat{\lambda}_{\mathrm{max}}$ on each flux surface. To do so, we need to define an objective function that depends on $\widehat{\lambda}_{\mathrm{max}}$ such that minimizing the objective function should allow us to achieve a stable equilibrium. Moreover, during optimization, once a flux surface is stabilized against the ideal ballooning mode, our objective function should ignore that particular surface. This would be useful as we do not want to penalize a stable equilibrium. To this end, we design the following ideal ballooning objective function
\begin{equation}
    f_{\mathrm{ball}} = \sum_{j=1}^{\mathrm{ns}} \mathrm{ReLU}(\widehat{\lambda}_{\mathrm{max},j} - \widehat{\lambda}_{\mathrm{th}, j}),
    \label{eqn:fiball}
\end{equation}
where $\mathrm{ns}$ is the total number of surfaces and
\[
    \mathrm{ReLU}(x)= 
\begin{cases}
    0,& \text{if } x\leq 0\\
    x,              & x> 0
\end{cases}
\]
\iffalse
\textcolor{red}{Adding $\lambda_{\mathrm{th}}$ makes things harder}
\fi
is the Rectified Linear Unit operator --- an operator that sets all the non-positive values to zero and $\widehat{\lambda}_{\mathrm{th},j}$ is the threshold below which we declare a surface ideal ballooning-stable. The value $\widehat{\lambda}_{j} = 0$ on the $j^{\mathrm{th}}$ surface implies marginal stability but we choose $\widehat{\lambda}_{\mathrm{th},j} = 0.0001$ to ensure that all the surfaces are slightly away from marginal ideal ballooning stability. An equilibrium is ideal ballooning stable if $f_{\mathrm{ball}} = 0$.

It is also important to prevent the optimizer from minimizing $f_{\mathrm{ball}}$ in a trivial manner. For example, for 2D equilibria, going to a larger aspect ratio value stabilizes the ideal ballooning mode. For 3D equilibria, the optimizer can sometimes reduce the minor radius, which, for a fixed toroidal flux, causes the magnetic field to increase. This lowers the overall $\beta$ and consequently the unstable curvature drive term. Similarly, if we allow rotational transforms to increase freely, the optimizer can sometimes create large gradients of $\iota $, generating large currents  which is suboptimal. To avoid achieving such trivial solutions and uninteresting equilibria, we add a combination of the following penalty terms to the optimizer:

\begin{enumerate}
    \item $f_{\mathrm{asp}} = (A - A_0)$ to penalize any deviation from the aspect ratio of the initial equilibrium
    \item $f_{\mathrm{minr}} = (a_{\mathrm{N}} - a_{\mathrm{N}0})$ to penalize any deviation from the minor radius of the initial equilibrium
    \item $f_{\langle B \rangle} = (\langle B \rangle - \langle B\rangle_0)$ to penalize any deviation of the volume-averaged magnetic field from its value in the initial equilibrium
    \item $f_{R_{\mathrm{c}}} = \int d\theta\,  \mathrm{ReLU}({-R_\mathrm{c}})$ where $R_{\mathrm{c}}$ is the radius of curvature of the boundary. This term penalizes any boundary shapes that are curved into the plasma
    \item $f_{\bar{\iota}} = (\bar{\iota} - \bar{\iota}_0)$ where $\bar{\iota}$ and $\bar{\iota}_0$ to penalize deviation of the mean rotational transform
    \item $\mathcal{G}_{\iota}  = \mathrm{ReLU}(\iota_{\mathrm{th}} - \iota)$ where $\iota_{\mathrm{th}}$ is some threshold value of $\iota$  
    \item $\mathcal{G}_{j_{\zeta}}  = \mathrm{ReLU}( j_{\zeta, \mathrm{th}} - j_{\zeta})$ where $j_{\zeta, \mathrm{th}}$ is some threshold value of the enclosed toroidal current \\
\end{enumerate}
Using the ballooning objective function~\eqref{eqn:fiball} and one or more of the penalty terms described above, we can get the overall objective function $\mathcal{F}$. Given a vector of input parameters $\boldsymbol{p}$, our goal is to solve
\begin{equation}
    \min_{\boldsymbol{p}} \mathcal{F}(\boldsymbol{p}), \qquad \textrm{s.t.} \quad f_{\mathrm{ball}} =  0.
\end{equation}
We achieve this with the help of the $\texttt{SIMSOPT}$~\citep{landreman2021simsopt} package. The implementation details of the optimization are described in the next subsection.

\subsection{Optimization with the $\mathtt{SIMSOPT}$ package}
\label{subsec:SIMSOPT}
In this subsection, we discuss the implementation-related details of an adjoint ballooning solver with the \texttt{SIMSOPT}~\citep{landreman2021simsopt} package. First, we briefly explain how an optimization problem can be solved using~\texttt{SIMSOPT}. Next, we go into the details of how we solve the ideal ballooning optimization and how the use of an adjoint method can speed up this process.

The \texttt{SIMSOPT} package is a optimization framework containing a suite of codes that, along with the $\texttt{VMEC}$ code, have been used to optimize 3D equilibria for various properties like energetic fast-particle confinement, quasisymmetry, simpler magnetic coil geometry, neoclassical transport, etc. The user specifies the input parameters (also referred to as Degrees of Freedom (Dofs)) and the objective function $\mathcal{F}$ and $\texttt{SIMSOPT}$ can perform a gradient-based or gradient-free nonlinear least squares optimization. 

As an example, let us construct an optimization problem to stabilize an equilibrium while penalizing the change in the aspect ratio and the minor radius of the boundary
\begin{equation}
    \mathcal{F} = f_{\mathrm{asp}}^2 + f_{\mathrm{minr}}^2 + f_{\mathrm{ball}}^2. 
\end{equation}
 To do so, we use gradient-based optimization in \texttt{SIMSOPT}, where one calculates $\partial \mathcal{F}/\partial \boldsymbol{p}$ to update the parameter vector at the $i^{\mathrm{th}}$ iteration, $\boldsymbol{p}_i$ as
\begin{equation}
    \boldsymbol{p}_{i+1} = f\left(\boldsymbol{p}_{i}, \frac{\partial \mathcal{F}}{\partial \boldsymbol{p}_i}\right).
\end{equation}
This is done until the optimizer reaches a local minima, i.e., a region in the parameter space where $\partial \mathcal{F}/\partial \boldsymbol{p} = 0$ or the relative change in the gradient is small enough. Typically, one has to evaluate the gradient of $\mathcal{F}$ hundreds of times during an optimization loop before finding a local minimum. 
In this study, evaluating $f_{\mathrm{ball}}$ is the most expensive step. Because the speed of the optimization is limited by the rate at which we can compute $f_{\mathrm{ball}}$, we have used an adjoint method to calculate $\widehat{\lambda}_{\mathrm{max}}$ which gives us $f_{\mathrm{ball}}$.

\section{Results}\label{sec:results}
In this section, we present the results of our study. In~\S\ref{subsec:2D-DIIID-result}, we compare the initial and optimized 2D axisymmetric equilibrium. In~\S\S\ref{subsec:3D-NCSX-result} and~\ref{subsec:3D-HBERG-result}, we do the same for the modified NCSX and modified Henneberg-QA equilibria, respectively. In addition, we also compare the values of relevant physical quantities in the initial and optimized equilibria.

\subsection{Stabilizing the DIII-D-like equilibrium}
\label{subsec:2D-DIIID-result}
\begin{figure}
    \centering
    {\includegraphics[width=0.68\textwidth, trim = 0mm 0mm 0mm 0mm, clip]{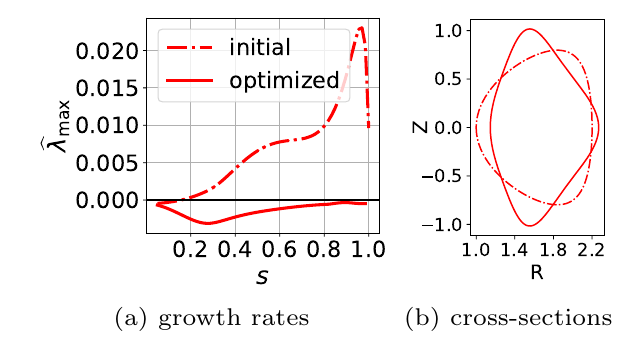}} 
    \caption{This figure shows ($\mathit{a}$) the maximum eigenvalue $\widehat{\lambda}_{\mathrm{max}}$ of the intial and optimized DIII-D-like equilibrium. The optimized equilibrium is stable. In ($\mathit{b}$), we present the boundary shape of the initial and final equilibria. Note the negative triangularity of the initial equilibrium and the positive triangularity of the optimized equilibrium.}
    \label{fig:DIIID-results}
\end{figure}

For the 2D axisymmetric case, we start with a high-$\beta$ equilibrium with a negative triangularity boundary. Due to axisymmetry, the 2D boundary does not depend on the toroidal angle $\zeta$, i.e., $n = 0$ in~\eqref{eqn:Fourier-boundary}. Therefore, the number of modes needed to specify a 2D boundary is much lower than that for a typical 3D boundary. In this problem, we pick the six largest Fourier modes as our Dofs: $\widehat{R}_{\rm{b}}(0, 1), \widehat{R}_{\rm{b}}(0, 2), \widehat{R}_{\rm{b}}(0, 3), \widehat{Z}_{\rm{b}}(0, 1), \widehat{Z}_{\rm{b}}(0, 3), \widehat{Z}_{\rm{b}}(0, 5)$. The full objective function is
\begin{equation}
    \mathcal{F} = f_{\mathrm{asp}}^2 + f_{R_{c}}^2 + f_{\mathrm{minr}}^2 + f_{\langle B \rangle}^2 + f_{\mathrm{ball}}^2 , 
\end{equation}
where all terms except $f_{\mathrm{ball}}$ are penalty terms to prevent the optimizer from producing a trivial solution. After this, we start with the negative triangularity equilibrium described in~\S\ref{subsec:2D-axisym-eqbm} and run \texttt{SIMSOPT} to find multiple equilibria that are stable against the ideal ballooning mode, i.e., equilibria with $f_{\mathrm{ball}} = 0$. We present one of the optimized equilibria in figure~\ref{fig:DIIID-results}. We also compare the values of equilibrium-dependent quantities in table~\ref{tab:Table-DIIID-eqbm-quant2}.

\begin{table}
  \begin{center}
\def~{\hphantom{0}}
  \begin{tabular}{l|ccccccccc}
    Eqbm. & $\beta_{\mathrm{ax}}(\%)$ &  $\langle \beta \rangle(\%)$  & $j_{\zeta} (\mathrm{M}A)$ & $\langle B \rangle$ (T)& $\bar{\iota}$ & $A$ & $a_{\mathrm{N}} (m)$ & $B_{\mathrm{N}}(T)$ & $\psi_{\mathrm{LCFS}}(T  m^2)$\\[3pt] \hline 
    Initial & $14.0$  & $7.6$  & $0.616 $ & $0.677$ & $0.568$ & $2.42$ & $0.68$ & $0.679$ & $1.0$ \\ \hline
    Optimized & $13.7$  & $7.3$  & $0.728 $ & $0.686$ & $0.568$ & $2.42$ & $0.68$ & $0.681$ & $1.0$ \\
  \end{tabular}
    \caption{This table shows a comparison between relevant physical quantities of the intial and optimized DIII-D equilibrium}
  \label{tab:Table-DIIID-eqbm-quant2}
  \end{center}
\end{table}

\iffalse
We were able to find a solution using \texttt{SIMSOPT}. \textcolor{red}{How long did the solver take}. We repeat the same process again, but without the adjoint method to find $\lambda_{max}$ and the solver takes~\textcolor{red}{X units of time}. 
\fi
 
We observe that the optimizer is moving toward a positive triangularity equilibrium, indicating that, for the similar values of the relevant parameters (given in table~\ref{tab:Table-DIIID-eqbm-quant2}) positive triangularity high-$\beta$ equilibria are more stable than their negative triangularity counterparts.
Our findings are consistent with recent observations by~\citet{davies2022kinetic} and~\citet{nelson2022prospects} that negative triangularity equilibria are more unstable against the ideal ballooning mode compared to positive triangularity ones. This behavior prevents the formation of a steep pressure gradient, which limits the operational beta value of negative triangularity equilibria. 

\subsection{Stabilizing the NCSX equilibrium}
\label{subsec:3D-NCSX-result}
\begin{figure}
    \centering
    {\includegraphics[width=1.03\textwidth, trim = 3mm 1mm 0mm 1mm, clip]{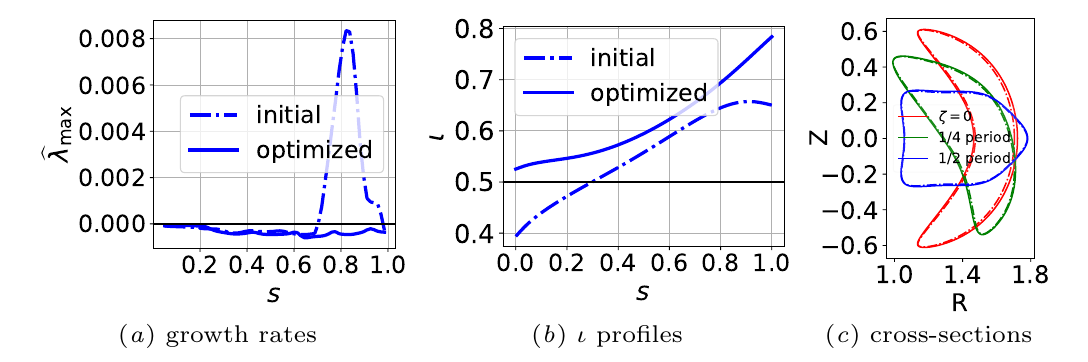}} 
    \caption{This figure shows ($\mathit{a}$) the maximum eigenvalue $\widehat{\lambda}_{\mathrm{max}}$ of the intial and optimized modified NCSX equilibrium. Similarly, in ($\mathit{b}$) we compare the rotational transform profiles of the initial and optimized equilibria. In ($\mathit{c}$), we present the boundary shape of the initial and final equilibria at three different value of the toroidal angle $\zeta$. The dotted curves correspond to the initial cross-sections whereas the solid curves are the final cross-sections.}
    \label{fig:NCSX-results}
\end{figure}

The first 3D equilibrium we optimize is an unstable NCSX equilibrium. Since the boundary has a 3D shape, we have to use both toroidal and poloidal modes to change its shape. For this demonstration, we choose $72$ boundary modes listed in table~\ref{tab:Table-NCSX-boundary} as Dofs: 
\begin{table}
  \begin{center}
\def~{\hphantom{0}}
  \begin{tabular}{lc}
    $\widehat{R}_{\mathrm{b}}(n, m)$ &  $\widehat{Z}_{\mathrm{b}}(n, m)$ \\[3pt]
    $([1,4], 0)$  & $([1, 4], 0)$ \\
    $([-3,3], 1)$  & $([-3, 3], 1)$ \\
    $([-3,3], 2)$  & $([-3, 3], 2)$ \\
    $([-2,2], 3)$  & $([-2, 2], 3)$ \\
    $([-2,2], 4)$  & $([-2, 2], 4)$ \\
    $([-2,2], 5)$  & $([-2, 2], 5)$ \\
    $([-1,1], 6)$  & $([-1, 1], 6)$ \\
  \end{tabular}
    \caption{This table provides the boundary shape Dofs for the NCSX case.}
  \label{tab:Table-NCSX-boundary}
  \end{center}
\end{table}
where $[i, j]$ denotes all integers between $i$ and $j$ (including $i$ and $j$). Additionally, we also provide the optimizer with the coefficients of the rotational transform profile $\iota(s)$. For this study, we have $6$ Dofs that determine the rotational transform profile. Therefore, we have a total of $78$ Dofs, much larger than the axisymmetric case. After choosing the Dofs, we choose the following general objective function
\begin{equation}
    \mathcal{F} = 0.5 f_{\mathrm{asp}}^2  + 0.5 f_{\mathrm{minr}}^2 + f_{\langle B \rangle}^2 + f_{\bar{\iota}}^2 + (70 f_{\mathrm{ball}})^2 , 
\end{equation}
We run \texttt{SIMSOPT} with this configuration to obtain multiple equilibria with $f_{\mathrm{ball}} = 0$. We have plotted a comparison of one of these equilibria with the initial equilibrium in~\ref{fig:NCSX-results}.

The optimizer stabilizes the equilibrium mostly through negative global magnetic shear $\hat{s} = -\iota d\iota/ds$. There is minimal change in the shape of the boundary. We also present a comparison of the important equilibrium-dependent quantities for the initial and optimized equilibria in table~\ref{tab:Table-NCSX-eqbm-quant2}.

\begin{table}
  \begin{center}
\def~{\hphantom{0}}
  \begin{tabular}{l|ccccccccc}
    Equilibrium & $\beta_{\mathrm{ax}}(\%)$ &  $\langle \beta \rangle(\%)$  & $j_{\zeta} (\mathrm{M}A)$ & $\langle B \rangle$ (T)& $\bar{\iota}$ & $A$ & $a_{\mathrm{N}} (m)$ & $B_{\mathrm{N}}(T)$ & $\psi_{\mathrm{LCFS}}(T  m^2)$\\[3pt] \hline 
    Initial & $9.3$  & $5.1$  & $0.174 $ & $1.596$ & $0.48$ & $4.36$ & $0.325$ & $1.54$ & $0.514$ \\ \hline
    Optimized & $8.2$  & $4.5$  & $0.215 $ & $1.694$ & $0.59$ & $4.48$ & $0.316$ & $1.63$ & $0.514$ \\
  \end{tabular}
    \caption{This table presents a comparison between relevant physical quantities of the intial and optimized NCSX equilibrium}
  \label{tab:Table-NCSX-eqbm-quant2}
  \end{center}
\end{table}

\subsection{Stabilizing the modified Henneberg-QA equilibrium}
\label{subsec:3D-HBERG-result}
As a final example, we present a modified modified Henneberg-QA equilibrium. In this example, we allow the boundary coefficients given in table~\ref{tab:Table-HBERG-boundary} to change freely.

\begin{table}
  \begin{center}
\def~{\hphantom{0}}
  \begin{tabular}{lc}
    $\widehat{R}_{\mathrm{b}}(n, m)$ &  $\widehat{Z}_{\mathrm{b}}(n, m)$ \\[3pt]
    $([1,5], 0)$  & $([1, 5], 0)$ \\
    $([-3,3], 1)$  & $([-3, 3], 1)$ \\
    $([-3,3], 2)$  & $([-3, 3], 2)$ \\
    $([-2,2], 3)$  & $([-2, 2], 3)$ \\
    $([-2,2], 4)$  & $([-2, 2], 4)$ \\
    $([-2,2], 5)$  & $([-2, 2], 5)$ \\
    $([-1,1], 6)$  & $([-1, 1], 6)$ \\
  \end{tabular}
    \caption{This table lists the boundary shape Dofs for the modified Henneberg-QA case.}
  \label{tab:Table-HBERG-boundary}
  \end{center}
\end{table}

\begin{figure}
    \centering
    {\includegraphics[width=1.0\textwidth, trim = 3mm 1mm 0mm 1mm, clip]{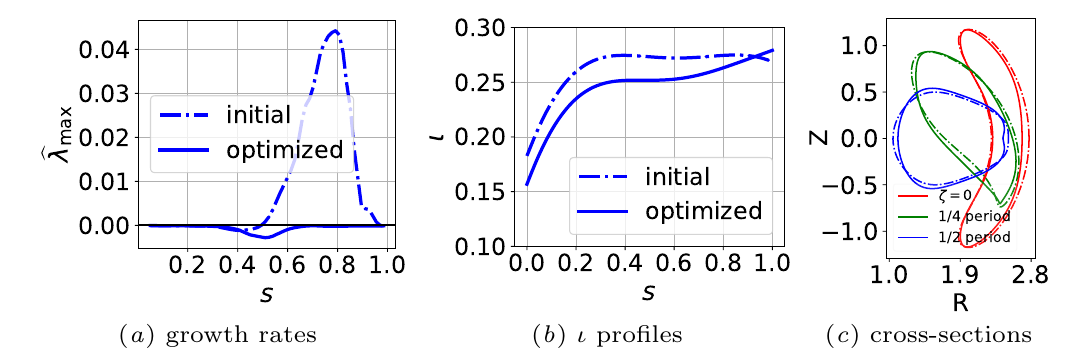}} 
    \caption{This figure shows ($\mathit{a}$) the maximum eigenvalue $\lambda_{\mathrm{max}}$ of the intial and optimized modified Henneberg-QA equilibrium. Similarly, in ($\mathit{b}$) we com    apre the rotational transform profiles of the intial and optimized equilibria. In ($\mathit{c}$), we present the boundary shape of the initial and final equilibria at three different positions  of the toroidal angle $\zeta$. The dotted curves correspond to the initial cross-  sections whereas the solid curves are the final cross-sections.}
    \label{fig:HBERG-QA-results}
\end{figure}

We also have seven coefficients that determine the rotational transform profile, giving us a total of $83$ Dofs. For this problem, we choose the following objective function
\begin{equation}
    \mathcal{F} = 0.1 f_{\mathrm{asp}}^2  + 0.1 f_{\mathrm{minr}}^2 + f_{\langle B \rangle}^2 + 10^{-8}\mathcal{G}_{j_{\zeta}} + 5\mathcal{G}_{\iota}^2 +  10 f_{\mathrm{ball}}^2 , 
\end{equation}
After choosing the Dofs and the objective function, we run \texttt{SIMSOPT} and obtain multiple stable equilibria. We compare one of the stable equilibria with the initial, unstable equilibrium in figure~\ref{fig:HBERG-QA-results} and the equilibrium-dependent quantities in table~\ref{tab:Table-HBERG-eqbm-quant2}.

\begin{table}
  \begin{center}
\def~{\hphantom{0}}
  \begin{tabular}{l|ccccccccc}
    Equilibrium & $\beta_{\mathrm{ax}}(\%)$ &  $\langle \beta \rangle(\%)$  & $j_{\zeta} (\mathrm{M}A)$ & $\langle B \rangle (T)$& $\bar{\iota}$ & $A$ & $a_{\mathrm{N}} (m)$ & $B_{\mathrm{N}} (T)$ & $\psi_{\mathrm{LCFS}}(T  m^2)$\\[3pt] \hline 
    Initial & $4.9$  & $2.65$  & $0.235 $ & $2.501$ & $0.263$ & $3.374$ & $0.601$ & $2.35$ & $2.677$ \\ \hline
    Optimized & $4.5$  & $2.47$  & $0.002 $ & $2.588$ & $0.245$ & $3.381$ & $0.592$ & $2.42$ & $2.677$ \\
  \end{tabular}
    \caption{This table shows a comparison between relevant physical quantities of the intial and optimized modified Henneberg-QA equilibrium}
  \label{tab:Table-HBERG-eqbm-quant2}
  \end{center}
\end{table}
We find that the equilibrium is stabilized by a combination of boundary shaping and rotational transform. The rotation transform creates negative magnetic shear in the outer core, whereas the boundary shape reduces the curvature in the ``bad''-curvature region.

\section{Summary and Conclusions}\label{sec:summary-and-conclusions}
We began this work by briefly explaining the various curvilinear coordinate systems that we used to fully define a general 3D ideal MHD equilibrium. In~\S\ref{sec:equilibrium}, we generated three different equilibria: one axisymmetric 2D and two 3D, and described the important associated physical quantities that could be used to penalize large deviations or trivial solutions generated by the optimizer. 

Upon generating the different equilibria, in~\S\ref{sec:ideal-balloning mode}, we provided a physical description and explain the numerical methods used to calculate the maximum eigenvalues on a given flux surface. Using these numerical methods, we evaluated the stability of all three chosen equilibria against the infinite-$n$ ideal ballooning mode. We also described the self-adjoint property of the ideal ballooning mode and its relation to the KBM.

Using the self-adjoint property explained in~\S\ref{sec:ideal-balloning mode}, we developed an adjoint method in~\S\ref{sec:Adjoint-method} and explained how to use it to speed up the calculation of the maximum ballooning eigenvalue $\lambda_{\mathrm{max}}$ on each surface. To demonstrate the efficiency and accuracy of the adjoint method, we also presented a comparison of gradients between an adjoint method and a finite difference scheme. We found that the adjoint method is up to four times faster than the finite-difference scheme.

In~\S\ref{sec:optimizn-details}, we describe the details of the overall optimization process and how we accomplish that using the~\texttt{SIMSOPT} code. After implementing the optimization, we presented the results in~\S\ref{sec:results}. We presented the specific details of the objective function and the Dofs for each equilibrium and stabilized the initial, ideal ballooning unstable equilibria. We briefly described the physical mechanism of the optimized equilibria that stabilize the ideal ballooning mode.

This work presents many avenues for future research. A key step forward is to extend our technique to include all equilibrium-dependent parameters $\widetilde{\boldsymbol{p}}$ as explained in the appendix~\ref{app:adjoint-iball-general}. One could also use the exact same method to optimize stellarators and tokamaks against low-$n$, unstable ideal MHD modes, as explained in the appendix~\ref{app:adjoint-iball-lown}. Since solving for low-$n$ ideal MHD codes is much more computationally expensive, the advantage of using an adjoint method would be even greater. Another possible direction would be to use an adjoint method to get derivatives of the ballooning growth rate with respect to the plasma shape. Finally, one could use the ideal balloon optimizer as a tool that could help optimize an equilibrium against KBMs. These optimizers could also help us look for equilibrium-dependent proxies for MHD or kinetic instabilities.

\textbf{Acknowledgements}: We thank Prof. David Bindel, Dr. Aaron Bader, and Dr. Ben Faber for helpful discussions and encouragement. One of the authors, R.G., thanks Rory Conlin for his valuable suggestions at the APS-DPP 2022 conference. This work was supported by the US Department of Energy, Office of Science, and Office of Fusion Energy Sciences under Award Numbers DE-SC0018429 and DE-FG02-93ER54197. This research used resources from the National Energy Research Scientific Computing Center, a DOE Office of Science User Facility, and the Stellar cluster at Princeton University. This material is based on work supported by the National Science Foundation Graduate Research Fellowship under Grant No. DGE-1650441.

\textbf{Declaration of interests}
The authors report that they do not have a conflict of interest.

\iffalse
\textbf{Data availability statement} \textcolor{green}{The GitHub/zenodo/Bitbucket? link containing all the equilibria in this study and the code is given here.}
\fi

\appendix

\section{The discretized ideal ballooning equation}
\label{app:ideal_ballooning_matrix}
After applying the boundary conditions $\widehat{X}_0 = \widehat{X}_N = 0$ to~\eqref{eqn:discrete-iball}, we can rewrite the ballooning equation as
\begin{equation}
    (A - \lambda \mathds{I}) \widehat{X} = 0,
\end{equation}
with $\mathds{I}$ being the identity matrix and the asymmetric tridiagonal matrix $A$ with the following form
\[\!\!
\begin{bmatrix}
\frac{\mathrm{g}_{1/2} +  \mathrm{g}_{3/2} - (\Delta \theta)^2 \mathrm{c}_{1}}{(\Delta \theta)^2 \mathrm{f}_{1}} \hspace*{-8mm} &  -\frac{\mathrm{g}_{3/2}}{(\Delta \theta)^2 \mathrm{f}_1} & \hspace*{-4mm}  0 &\hspace*{-3mm} 0 &  \hspace*{-4mm}\ldots & \hspace*{-6mm} 0 & \hspace*{-10mm} 0\\
-\frac{\mathrm{g}_{3/2}}{(\Delta \theta)^2 \mathrm{f}_2} \hspace*{-7mm} & \frac{\mathrm{g}_{3/2}  + \mathrm{g}_{5/2} - (\Delta \theta)^2 \mathrm{c}_{2}}{(\Delta \theta)^2 \mathrm{f}_{2}}  & \hspace*{-2mm} -\frac{\mathrm{g}_{5/2}}{(\Delta \theta)^2 \mathrm{f}_2} & \hspace*{-3mm} 0  & \hspace*{-4mm} \ldots & \hspace*{-6mm}  0 & \hspace*{-10mm} 0\\
& & & \ddots\\
& &  & & \ddots\\
0 & 0 &\hspace*{-2mm} 0 & \hspace*{-2mm} 0 & \hspace*{-4mm} \ldots &  \hspace*{-4mm}\frac{\mathrm{g}_{N\!-\!5/2}  + \mathrm{g}_{N\!-\!3/2} - (\Delta \theta)^2 \mathrm{c}_{N\!-\!2}}{h^2 \mathrm{f}_{N\!-\!2}}  & \hspace*{-10mm}  -\frac{\mathrm{g}_{N\!-\!3/2}}{(\Delta \theta)^2 \mathrm{f}_{N\!-\!2}}\\
0 & 0 &\hspace*{-2mm} 0 & \hspace*{-2mm} 0 & \hspace*{-4mm} \ldots &  \hspace*{-6mm} -\frac{\mathrm{g}_{N\!-\!3/2}}{(\Delta \theta)^2 \mathrm{f}_{N\!-\!1}} & \hspace*{-10mm}  \frac{\mathrm{g}_{N\!-\!3/2}  + \mathrm{g}_{N\!-\!1/2} - (\Delta \theta)^2\! \mathrm{c}_{N\!-\!1}}{(\Delta \theta)^2 \mathrm{f}_{N\!-\!1}} \\
\end{bmatrix}\!\!\!\! \! \!\!
\]

\section{Details of the adjoint ideal ballooning calculation}
\label{app:adjoint-iball-math}
In this appendix, we will derive~\eqref{eqn:dlamdp-main} starting with~\eqref{eqn:iball-condensed}. To do that, we will use the self-adjoint property of ideal MHD given in~\eqref{eqn:self-adjoint-property1} as well as the Dirichlet boundary conditions satisfied by the eigenfunction, 
\begin{equation}
\widehat{X}(\theta = \pm \theta_{\mathrm{b}}) = \widehat{X}^{*}(\theta = \pm \theta_{\mathrm{b}}) = 0.
\label{eqn:DBC}
\end{equation}
Defining $\boldsymbol{p} = (\widetilde{\boldsymbol{p}}, \widehat{\boldsymbol{p}})$ as the union of all the parameters of the problem, we start by taking the derivative $\partial /\partial \boldsymbol{p}$ of~\eqref{eqn:iball-condensed},
\begin{equation}
    \frac{d}{d\theta} \frac{\partial \mathrm{g}}{\partial \boldsymbol{p}} \frac{d \widehat{X}}{d\theta}  + \frac{d}{d\theta} \mathrm{g} \frac{d }{d\theta}\frac{\partial \widehat{X}}{\partial \boldsymbol{p}}  + \frac{\partial \mathrm{c}}{\partial \boldsymbol{p}} \widehat{X}  + c\frac{\partial \widehat{X}}{\partial \boldsymbol{p}} = \frac{\partial \widehat{\lambda}}{\partial \boldsymbol{p}} \mathrm{f} \widehat{X} + \widehat{\lambda} \mathrm{f} \frac{\partial \widehat{X}}{\partial \boldsymbol{p}} + \widehat{\lambda} \widehat{X} \frac{\partial \mathrm{f}}{\partial \boldsymbol{p}}.
\end{equation}
Multiplying with $\widehat{X}^{*}$ on both sides, integrating throughout the domain, 
\begin{equation}
\begin{split}
     \int_{-\theta_{\mathrm{b}}}^{\theta_{\mathrm{b}}} d\theta\,   \widehat{X}^{*}\frac{d}{d\theta} \frac{\partial \mathrm{g}}{\partial \boldsymbol{p}} \frac{d \widehat{X}}{d\theta} &+ \int_{-\theta_{\mathrm{b}}}^{\theta_{\mathrm{b}}}  \widehat{X}^{*}\frac{d}{d\theta} \mathrm{g} \frac{d}{d\theta}\frac{\partial \widehat{X}}{\partial \boldsymbol{p}}  + \int_{-\theta_{\mathrm{b}}}^{\theta_{\mathrm{b}}} d\theta\, \frac{\partial \mathrm{c}}{\partial \boldsymbol{p}} \lvert \widehat{X} \rvert^2 + \int_{-\theta_{\mathrm{b}}}^{\theta_{\mathrm{b}}} d\theta \, \mathrm{c} \widehat{X}^{*} \frac{\partial \widehat{X}}{\partial \boldsymbol{p}}\\
    =&\ \frac{\partial \widehat{\lambda}}{\partial \boldsymbol{p}} \int_{-\theta_{\mathrm{b}}}^{\theta_{\mathrm{b}}} d\theta \, \mathrm{f} \lvert \widehat{X} \rvert^2 + \widehat{\lambda}  \int_{-\theta_{\mathrm{b}}}^{\theta_{\mathrm{b}}} d\theta \, \frac{\partial \mathrm{f}}{\partial \boldsymbol{p}} \lvert \widehat{X}\rvert^2 +   \widehat{\lambda} \int_{-\theta_{\mathrm{b}}}^{\theta_{\mathrm{b}}} d\theta \, \mathrm{f} \widehat{X}^{*} \frac{\partial \widehat{X}}{\partial \boldsymbol{p}}. 
\end{split}
\label{eqn:big-iball-integral}
\end{equation}
Using integration by parts,~\eqref{eqn:DBC} and rearranging~\eqref{eqn:big-iball-integral}, we can write
\begin{equation}
\begin{split}
\int_{-\theta_{\mathrm{b}}}^{\theta_{\mathrm{b}}} d\theta \bigg(\dfrac{\partial \mathrm{c}}{\partial \boldsymbol{p}}\lvert \widehat{X}\rvert^2 - \dfrac{\partial g}{\partial \boldsymbol{p}}\Big\lvert \dfrac{d\widehat{X}}{d\theta}\Big\rvert^2 -&  \widehat{\lambda} \dfrac{\partial \mathrm{f}}{\partial \boldsymbol{p}} \lvert \widehat{X}\rvert^2 \bigg)- \frac{\partial \widehat{\lambda}}{\partial \boldsymbol{p}} \int_{-\theta_{\mathrm{b}}}^{\theta_{\mathrm{b}}} d\theta \, \mathrm{f} \lvert \widehat{X} \rvert^2\\
=&  \int_{-\theta_{\mathrm{b}}}^{\theta_{\mathrm{b}}} d\theta \bigg( -\frac{d}{d\theta} \mathrm{g} \frac{d\widehat{X}^{*}}{d\theta} - \mathrm{c} \widehat{X}^{*} + \widehat{\lambda} \mathrm{f} \widehat{X}^{*} \bigg) \frac{\partial \widehat{X}}{\partial \boldsymbol{p}}.
\end{split}
\label{eqn:big-iball-integral2}
\end{equation}
Due to the self-adjoint property of ideal MHD, the right side of~\eqref{eqn:big-iball-integral2} is zero. The rest of the equation can be arranged so that
\begin{equation}
    \frac{\partial \widehat{\lambda}}{\partial \boldsymbol{p}} = \frac{\bigintss_{-\theta_{\mathrm{b}}}^{\theta_{\mathrm{b}}} d\theta \left(\dfrac{\partial \mathrm{c}}{\partial \boldsymbol{p}}\lvert \widehat{X}\rvert^2 - \dfrac{\partial \mathrm{g}}{\partial \boldsymbol{p}}\Big\lvert \dfrac{d\widehat{X}}{d\theta}\Big\rvert^2 -  \widehat{\lambda} \dfrac{\partial \mathrm{f}}{\partial \boldsymbol{p}} \lvert \widehat{X}\rvert^2 \right)}{\int_{-\theta_{\mathrm{b}}}^{\theta_{\mathrm{b} }} d\theta\,  \mathrm{f}\lvert \widehat{X}\rvert^2}.
\end{equation}
Therefore, to calculate $\partial \widehat{\lambda}/\partial \boldsymbol{p}$, we only need the gradients of the geometric coefficients $\mathrm{g}, \mathrm{c}$, $\mathrm{f}$, the eigenfunction $\widehat{X}$, and the eigenvalue $\widehat{\lambda}$ of the ballooning equation; we have to solve the ideal ballooning equation only once. This speeds up the optimization loop, as it is much faster to obtain the gradient of the geometric coefficients than to solve the ballooning equation multiple times.

\subsection{Extending our adjoint-based technique to equilibrium-dependent degrees of freedom}
\label{app:adjoint-iball-general}
In this paper, we have used an adjoint method to find the maximum eigenvalue $\lambda_{\rm{max}}$ on a flux surface. It is possible to extend our method to minimize $f_{\rm{ball}}$ under the appropriate conditions. We define the problem and find the pertinent conditions in this appendix. We want to find
\begin{equation}
    \min f_{\rm{ball}}(\widehat{\lambda}_{\mathrm{max}}, \widetilde{\boldsymbol{p}}, \widehat{\boldsymbol{p}}), \qquad \textrm{s.t.} \quad \mathcal{G}(\widehat{\lambda}, \widehat{X}, \widetilde{\boldsymbol{p}}, \widehat{\boldsymbol{p}}) \equiv \mathcal{L}\widehat{X} - \widehat{\lambda} \widehat{X} =  0. 
\end{equation}
where all symbols are defined in~\S\ref{sec:Adjoint-method} and the ballooning objective function $f_{\rm{ball}}$ is defined in~\eqref{eqn:fiball}. To minimize $f_{\rm{ball}}$ with respect to the equilibrium parameters, we need 
\begin{equation}
    \frac{d f_{\rm{ball}}}{d \widetilde{\boldsymbol{p}}} = \frac{\partial f_{\rm{ball}}}{\partial \widehat{\lambda}_{\rm{max}}}\bigg\lvert_{\widetilde{\boldsymbol{p}}, \widehat{\boldsymbol{p}}} \frac{\partial \widehat{\lambda}_{\rm{max}}}{\partial \widetilde{\boldsymbol{p}}}\bigg\lvert_{\widehat{\boldsymbol{p}}}+ \frac{\partial f_{\rm{ball}}}{\partial \widehat{\boldsymbol{p}}}\bigg\lvert_{\widehat{\lambda}_{\rm{max}}, \widetilde{\boldsymbol{p}}} \frac{\partial \widehat{\boldsymbol{p}}}{\partial \widetilde{\boldsymbol{p}}}\bigg\lvert_{\widehat{\lambda}_{\rm{max}}} + \frac{\partial f_{\rm{ball}}}{\partial \widetilde{\boldsymbol{p}}}\bigg\lvert_{\widehat{\lambda}_{\rm{max}}, \widehat{\boldsymbol{p}}}.
    \label{eqn:dfdp}
\end{equation}
The most expensive term to calculate in~\eqref{eqn:dfdp} is the gradient of the eigenvalue $\lambda$. To obtain that, we take the derivative of the operator $\mathcal{G}$ with respect to $\lambda$,
\begin{equation}
    -\frac{\partial \mathcal{G}}{\partial \widehat{\lambda}}\frac{\partial \widehat{\lambda}}{\partial \widetilde{\boldsymbol{p}}} =  \frac{\partial \mathcal{G}}{\partial \widehat{\lambda}}\frac{\partial \widehat{\lambda}}{\partial \widehat{\boldsymbol{p}}} + \frac{\partial \mathcal{G}}{\partial \widehat{X}} \frac{\partial \widehat{X}}{\partial \widehat{\boldsymbol{p}}} + \frac{\partial \mathcal{G}}{\partial \widehat{X}} \frac{\partial \widehat{X}}{\partial \widetilde{\boldsymbol{p}}} + \frac{\partial \mathcal{G}}{\partial \widehat{\boldsymbol{p}}} + \frac{\partial \mathcal{G}}{\partial \widetilde{\boldsymbol{p}}}.
    \label{eqn:dGdp2}
\end{equation}
We also express $\widehat{\lambda}$ around a point $\boldsymbol{p}_0 = (\widetilde{\boldsymbol{p}}_0, \widehat{\boldsymbol{p}}_0)$ in the state space as a Taylor series,
\begin{equation}
    \widehat{\lambda} = \widehat{\lambda}(\boldsymbol{p}_0) + \frac{\partial \widehat{\lambda}}{\partial \widetilde{\boldsymbol{p}}} \cdot \delta \widetilde{\boldsymbol{p}} + \frac{\partial \widehat{\lambda}}{\partial \widehat{\boldsymbol{p}}} \cdot \delta \widehat{\boldsymbol{p}}  +  \frac{\partial}{{\partial \widetilde{\boldsymbol{p}}} }\frac{\partial \widehat{\lambda}}{\partial \widetilde{\boldsymbol{p}}} \boldsymbol{:} \delta \widetilde{\boldsymbol{p}} \delta \widetilde{\boldsymbol{p}} +  \frac{\partial}{{\partial \widehat{\boldsymbol{p}}} }\frac{\partial \widehat{\lambda}}{\partial \widehat{\boldsymbol{p}}} \boldsymbol{:} \delta \widehat{\boldsymbol{p}} \delta \widehat{\boldsymbol{p}} + \textit{O}(|\delta\boldsymbol{p}|^3),
    \label{eqn:lambda_Taylor1}
\end{equation}
and assuming that the optimizer takes a step size $|\delta\boldsymbol{p}|$ that is smaller than the radius of convergence of Taylor series~\eqref{eqn:lambda_Taylor1},  
\begin{equation}
\bigg \lvert\frac{\partial}{{\partial \widetilde{\boldsymbol{p}}} }\frac{\partial \widehat{\lambda}}{\partial \widetilde{\boldsymbol{p}}} \boldsymbol{:} \delta \widetilde{\boldsymbol{p}} \delta \widetilde{\boldsymbol{p}} \bigg\rvert \ll \bigg\lvert \frac{\partial \widehat{\lambda}}{\partial \widetilde{\boldsymbol{p}}} \cdot \delta \widetilde{\boldsymbol{p}}  \bigg\rvert,\quad  \bigg \lvert\frac{\partial}{{\partial \widehat{\boldsymbol{p}}} }\frac{\partial \widehat{\lambda}}{\partial \widehat{\boldsymbol{p}}} \boldsymbol{:} \delta \widehat{\boldsymbol{p}} \delta \widehat{\boldsymbol{p}} \bigg\rvert \ll \bigg\lvert \frac{\partial \widehat{\lambda}}{\partial \widehat{\boldsymbol{p}}} \cdot \delta \widehat{\boldsymbol{p}}  \bigg\rvert.
    \label{eqn:lambda_Taylor2}
\end{equation}
 Using~\eqref{eqn:lambda_Taylor1} and~\eqref{eqn:lambda_Taylor2}
\begin{equation}
    \frac{\partial \widehat{\lambda}}{\partial \widetilde{\boldsymbol{p}}} = \frac{\partial \widehat{\lambda}}{\partial \widehat{\boldsymbol{p}}} +  \frac{\partial \widehat{\lambda}}{\partial \widetilde{\boldsymbol{p}}} \cdot \frac{\partial \widetilde{\boldsymbol{p}}}{\partial \widehat{\boldsymbol{p}}}.
\end{equation}
Next, we use the fact that $\partial \widehat{\lambda}/\partial \widehat{\boldsymbol{p}} = 0$ at $\widehat{\lambda} = \widehat{\lambda}_{\mathrm{max}}$, and that our choice of $f_{\rm{ball}}$ only explicitly depends on $\widehat{\lambda}_{\rm{max}}$. Using the explicit form of the linear operator from~\eqref{eqn:ballooning-equation}, we multiply equation~\eqref{eqn:dGdp2} by $X^{*}$ and integrate throughout the domain, to rewrite~\eqref{eqn:dfdp} as
\begin{equation}
    \frac{d f_{\rm{ball}}}{d\boldsymbol{p}} = \sum_{j=1}^{\rm{ns}} \mathrm{ReLU}^{'}(\widehat{\lambda}_{\mathrm{max},j})\frac{\bigintss_{-\theta_{\mathrm{b}}}^{\theta_{\mathrm{b}}} d\theta \left(\dfrac{\partial \mathrm{c}}{\partial \boldsymbol{p}} \lvert \widehat{X}\rvert^2 - \dfrac{\partial g}{\partial \boldsymbol{p}} \Big\lvert \dfrac{d\widehat{X}}{d\theta}\Big\rvert^2 -  \widehat{\lambda} \dfrac{\partial \mathrm{f}}{\partial \boldsymbol{p}} \lvert \widehat{X}\rvert^2 \right)}{\int_{-\theta_{\mathrm{b}}}^{\theta_{\mathrm{b} }} d\theta\,  \mathrm{f}\lvert \widehat{X}\rvert^2}, 
    \label{eqn:adjoint-relation-wrt-eqbm}
\end{equation}
where $\boldsymbol{p} = (\widetilde{\boldsymbol{p}}, \widehat{\boldsymbol{p}})$ is the union of all the parameters of the problem and $\mathrm{ReLU}^{'}$\footnote{The derivative of the $\mathrm{ReLU}$ operator is not well-defined at $x=0$. We may have to replace it with an activation function that is continuous with a well-defined derivative. For example, we could use the logistic function $1/(1 + e^{-c x})$ with a large positive real number $c$.} is the derivative of the $\mathrm{ReLU}$ operator such that
\[
    \mathrm{ReLU}^{'}(x)= 
\begin{cases}
    0,& \text{if } x < 0\\
    1,              & x> 0
\end{cases}
\]
Calculating the derivative of the geometric coefficients $\mathrm{g}, \rm{c}$, and $\rm{f}$, with respect to the equilibrium parameter vector $\widetilde{\boldsymbol{p}}$ is not straightforward in \texttt{VMEC} and may lack the requisite accuracy for an adjoint method to work. However, an equilibrium solver like \texttt{DESC}~\citep{dudt2020desc} that is designed to calculate these gradients along with the geometric coefficients accurately may enable us to utilize the full potential of this adjoint-based method. Since the speed up obtained with an adjoint method is 
 linearly proportional to the length of the vector $\widetilde{\boldsymbol{p}}$, using~\eqref{eqn:adjoint-relation-wrt-eqbm} we can, in principle, speed up the calculation of $d f_{\rm{ball}}/d\widetilde{\boldsymbol{p}}$ by an order of magnitude for 2D axisymmetric equilibria and by two orders or magnitude for 3D equilibria.

\subsection{Extending our adjoint technique to low-$n$, ideal MHD solvers}
\label{app:adjoint-iball-lown}
Note that this process can be applied to any ideal MHD eigenvalue solver. For fluctuations that are not confined to a flux surface, one can solve for a fluctuation of the form
\begin{equation}
    X = \sum_{m,n} \widehat{X}_{m, n}(\psi) e^{i(m\theta-n\zeta)}
\end{equation}
where $X = \{X_{\psi}, X_{\alpha}\}$ are components of fluctuation $X$ perpendicular to the equilibrium magnetic field line, and $m$ and $n$ are the poloidal and toroidal mode numbers, respectively. We solve for $\widehat{X}(\psi)$, using codes such as $\texttt{ELITE}$ and $\texttt{GATO}$~\citep{bernard1981gato} for axisymmetric equilibria and $\texttt{CAS-3D}$~\citep{schwab1993ideal} or $\texttt{TERPSICHORE}$~\citep{anderson1990terpsichore} for 3D equilibria. For $\texttt{GATO}$ and $\texttt{CAS-3D}$ and $\texttt{TERPSICHORE}$, the ideal MHD energy principle is used to solve the matrix equation
\begin{equation}
    \mathcal{A} X = \lambda \mathcal{B} X,
\end{equation}
where $\mathcal{A}$ and $\mathcal{B}$ are real symmetric matrices. Currently, solving such an equation using these codes takes at least a few minutes for each mode. For such a problem, we can repeat the process explained at the beginning of this appendix to obtain the gradient,
\begin{equation}
    \frac{\partial \lambda}{\partial \boldsymbol{p}} = X^{\mathrm{T}}\left( \frac{\partial \mathcal{A}}{\partial \boldsymbol{p}} - \lambda \frac{\partial \mathcal{B}}{\partial \boldsymbol{p}}\right)X \Big/ {X^{\mathrm{T}}\mathcal{B} X}, 
    \label{eqn:dlam-dp-low-n}
\end{equation}
for all modes. Equation~\eqref{eqn:dlam-dp-low-n} is similar to the Hellman-Feynman theorem~\citep{hellmann1933rolle, feynman1939forces}. For axisymmetric equilibria, combining gradient information with fast equilibrium solvers such as $\texttt{EFIT}$~\citep{lao1985reconstruction} can help mitigate real-time disruption. One could also couple this adjoint approach with an optimizer to find low-$n$, ideal MHD stable equilibria.  

\iffalse
Some of these codes use a variational method to write the equation,
\begin{equation}
    \mathsf{F}[\boldsymbol{\xi}] = \lambda \, \boldsymbol{\xi},
\end{equation}
where $\mathsf{F}$ is the ideal MHD force operator~\citep{IdealMHD} as
\begin{equation}
    \lambda = \frac{\int d^3 \boldsymbol{r}\,  \boldsymbol{\xi} \cdot \mathsf{F}[\boldsymbol{\xi}]}{\int d^3\boldsymbol{r}  \lvert \xi \rvert^2} 
\end{equation}
For such a system, using the self-adjoint property of ideal MHD and the same techniques as used for the ideal ballooning equation, we can obtain the following result
\begin{equation}
    \frac{\partial \lambda}{\partial \boldsymbol{p}} = \frac{\int d^3 \boldsymbol{r}\,  \boldsymbol{\xi} \cdot \frac{\partial \mathsf{F}}{\partial \boldsymbol{p}}[\boldsymbol{\xi}]}{\int d^3\boldsymbol{r}  \lvert \xi \rvert^2} 
\end{equation}
where the partial derivative $\partial/\partial \boldsymbol{p}$ is applied to all equilibrium-dependent quantities in the functional $\mathsf{F}$.
\fi
\bibliographystyle{jpp}

\bibliography{jpp-instructions}

\end{document}